\documentclass[format=acmsmall]{acmart}

\usepackage{acm-ec-19}
\usepackage{tikz} % SW: Added for plots
\usepackage{subfiles}
\usepackage[toc,page]{appendix}
\usepackage{url}            % simple URL typesetting
\usepackage{booktabs}       % professional-quality tables
\usepackage{nicefrac}       % compact symbols for 1/2, etc.
\usepackage{microtype}      % microtypography
\usepackage{amsmath,amssymb,amsfonts}

\usepackage{amsthm} % The amsthm package provides extended theorem environments
\usepackage{float}
\usepackage{sgame, tikz} % Game theory packages
\usepackage{caption}
\usepackage{algorithm,algpseudocode}
\usepackage{makecell}
 \usepackage{multirow}
 \usepackage{graphicx}
\theoremstyle{definition}
\newtheorem{finding}{Finding}

\usepackage{xspace}
\newcommand{\OMIT}[1]{}
\newcommand{\xhdr}[1]{\vspace{1mm} \noindent{\bf #1}}
\newcommand{\ie}{{\em i.e.,~\xspace}}
\newcommand{\eg}{{\em e.g.,~\xspace}}

\newcommand{\term}[1]{\ensuremath{\mathtt{#1}}\xspace}
\newcommand{\TS}{\term{TS}}
\newcommand{\DEG}{\term{DEG}}
\newcommand{\DG}{\term{DG}}
\newcommand{\Beta}{\term{Beta}} % for Beta distribution
\newcommand{\Eeog}{\term{EoG}} % shorthand for "effective end of game"

\newcommand{\eps}{\varepsilon}

\newcommand{\MRV}{mean reward vector\xspace} % mean reward vector
\newcommand{\MRVs}{mean reward vectors\xspace} % mean reward vector

\newcommand{\HMR}{\term{HMR}}
\newcommand{\HM}{\term{HM}}
\usepackage[suppress]{color-edits}
\addauthor{as}{red}    % as for Alex
\addauthor{ga}{blue}  % ga for Guy
\addauthor{sw}{orange} % sw for Steven
\usepackage{booktabs} % For formal tables

\setcopyright{rightsretained}
\begin{document}
\title[The Perils of Exploration under Competition: A Computational Modeling Approach]
{The Perils of Exploration under Competition: \\ A Computational Modeling Approach}

 \author{Guy Aridor}
 \affiliation{ \institution{Columbia University}}
 \author{Kevin Liu}
 \affiliation{\institution{Columbia University}}
 \author{Aleksandrs Slivkins}
 \affiliation{\institution{Microsoft Research, New York City}}
 \author{Zhiwei Steven Wu}
 \affiliation{\institution{University of Minnesota - Twin Cities}}
 
 \authorsaddresses{%
  Authors' addresses: Guy Aridor, Columbia University, New York, NY, g.aridor@columbia.edu; Kevin Liu, Columbia University, New York, NY, kevin.liu@columbia.edu; Aleksandrs Slivkins, Microsoft Research, New York, NY, slivkins@microsoft.com; Zhiwei Steven Wu, University of Minnesota - Twin Cities, Minneapolis, MN, zsw@umn.edu}

\begin{abstract}
  We empirically study the interplay between \textit{exploration} and
  \textit{competition}. Systems that learn from interactions with
  users often engage in \emph{exploration}: making potentially
  suboptimal decisions in order to acquire new information for future
  decisions. However, when multiple systems are competing for
    the same market of users, exploration may hurt a system's
    reputation in the near term, with adverse competitive effects. In particular, a system may enter a ``death spiral", when the short-term reputation cost decreases
    the number of users for the system to learn from, which degrades the
    system's performance relative to competition and further decreases
    the market share.

We ask whether better exploration algorithms are incentivized under competition. We run extensive numerical experiments in a stylized duopoly model in which two firms deploy multi-armed bandit algorithms and compete for myopic users.  We find that duopoly and monopoly tend to favor a primitive ``greedy algorithm" that does not explore and leads to low consumer welfare, whereas a temporary monopoly (a duopoly with an early entrant) may incentivize better bandit algorithms and lead to higher consumer welfare. Our findings shed light on the first-mover advantage in the digital economy by exploring the role that data can play as a barrier to entry in online markets.
\end{abstract}

\maketitle

\section{Introduction}\label{sec:intro}

Many modern online platforms simultaneously compete for users as well as learn from the users they manage to attract. This creates a tension between \textit{exploration} and \textit{competition}: firms experiment with potentially sub-optimal options for the sake of gaining information to make better decisions tomorrow, while they need to incentivize consumers to select them over their competitors today. For instance, Google Search and Bing compete for users in the search engine market yet at the same time need to experiment with their search and ranking algorithms to learn what works best. Similar exploration vs. competition tension arises in other application domains: recommendation systems, news and entertainment websites, online commerce, and so forth.

Online platforms routinely deploy A/B tests, and are increasingly adopting  more sophisticated exploration methodologies based on \emph{multi-armed bandits}, a well-known framework for making decisions under uncertainty and trading off exploration and exploitation (making good near-term decisions). While deploying ``better" learning algorithms for exploration would improve performance, this is not necessarily beneficial under competition, even putting aside the deployment/maintenance costs. In particular, excessive experimentation may hurt a platform's reputation and decrease its market share in the near term. This would leave the learning algorithm with less users to learn from, which may further degrade the platform's performance relative to competitors who keep learning and improving from \emph{their} users, and so forth. Taken to the extreme, such dynamics may create a ``death spiral" effect when the vast majority of customers eventually switch to competitors.

In this paper, we ask how the interplay of exploration and competition affects platforms' incentives. While some bandit algorithms are traditionally considered ``better" than others in the literature, {\bf\em does competition incentivize the adoption of the better algorithms}? How is this affected by the intensity of competition? We investigate these issues via extensive numerical experiments in a stylized duopoly model.

\xhdr{Our model.} We consider two firms that compete for users and simultaneously learn from them. Each firm commits to a multi-armed bandit algorithm, and \emph{explores} according to this algorithm. Users select between the two firms based on the current reputation score: rewards from the firm's algorithm, averaged over a recent time window. Each firm's objective is to maximize its market share (the fraction of users choosing this firm).

We consider a \emph{permanent duopoly} in which both firms start at the same time, as well as \emph{temporary monopoly}: a duopoly with a first-mover. Accordingly, the intensity of competition in the model varies from ``permanent monopoly" (just one firm) to ``incumbent" (the first-mover in temporary monopoly) to permanent duopoly to ``entrant" (late-arriver in temporary monopoly).%
\footnote{\asedit{We consider the ``permanent monopoly" scenario for comparison only, without presenting any findings. We just assume that a monopolist chooses
the greedy algorithm, because it is easier to deploy in practice. Implicitly, users have no ``outside option": the service provided is an improvement over not having it (and therefore the monopolist is not incentivized to deploy better learning algorithms). This is plausible with free ad-supported platforms such as Yelp or Google.}}

We focus on three classes of bandit algorithms, ranging from more primitive to more sophisticated: \emph{greedy algorithms} that do not explicitly explore, algorithms that separate exploration and exploitation, and algorithms that combine the two. We know from prior work that in the absence of competition,  ``better" algorithms are better in the long run, but could be worse initially.

\xhdr{Main findings.}
We find that in the permanent duopoly, competition incentivizes firms to choose the ``greedy algorithm", and even more so if the firm is a late arriver in a market. This algorithm also prevails under monopoly, simply because it tends to be easier to deploy. Whereas the incumbent in the temporary monopoly is incentivized to deploy a more advanced exploration algorithm. As a result, consumer welfare is highest under temporary monopoly. We find strong evidence of the ``death spiral" effect mentioned earlier; this effect is strongest under permanent duopoly.

Interpreting the adoption of better algorithms as ``innovation", our
findings can be framed in terms of the ``inverted-U relationship"
between competition and innovation (see Figure~\ref{fig:inverted-U}).
This is a well-established concept in the economics literature, dating
  back to \cite{Schumpeter-42}, whereby too
  little or too much competition is bad for innovation, but
  intermediate levels of competition tend to be better. However, the
``inverted-U relationship'' is driven by different aspects in our
model than the ones in the existing literature in economics. In our
case, the barriers for innovations arise entirely from the
reputational consequences of exploration in competition, as opposed to
the R\&D costs (which is the more standard cause in prior work).

\tikzstyle{level 1}=[level distance=3.5cm, sibling distance=4.0cm]
\tikzstyle{level 2}=[level distance=3.5cm, sibling distance=2cm]
\tikzstyle{below} = [align=center]

\begin{figure}[t]
\begin{center}
\begin{tikzpicture}
      \draw[->] (-.5,0) -- (7,0) node[right] {};
      \draw[->] (0,-.5) -- (0,3) node[above] {Better algorithms};
      \draw[scale=0.6,domain=0.7:9.8,smooth,variable=\x,blue, line width=0.3mm] plot ({\x},{4.5 - 0.18 * (\x - 5.25)^2});
     \node[below] at (0.7, -0.22) {\footnotesize monopoly};
     \node[below] at (3, -0.2) {\footnotesize incumbent};
     \node[below] at (4.5, -0.22) {\footnotesize duopoly};
     \node[below] at (6, -0.2) {\footnotesize entrant};
 \end{tikzpicture}
 \caption{A stylized ``inverted-U relationship" between strength of competition and ``level of innovation".}
\label{fig:inverted-U}
\end{center}
\end{figure}
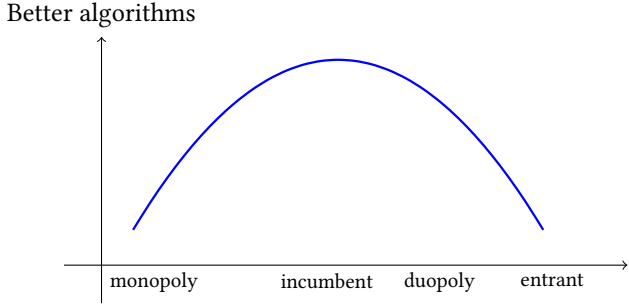

\xhdr{Additional findings.}
We investigate the ``first-mover advantage" phenomenon in more detail. Being first in the market gives free data to learn from (a ``data advantage") as well as a more definite, and possibly better reputation compared to an entrant (a ``reputation advantage"). We run additional experiments so as to isolate and compare these two effects. We find that either effect alone leads to a significant advantage under competition. The data advantage is larger than reputation advantage when the incumbent commits to a more advanced bandit algorithm.

Data advantage is significant from the anti-trust perspective, as a possible barrier to entry. We find that even a small amount ``data advantage" gets amplified under competition, causing a large difference in eventual market shares. This observation runs contrary to prior work  \cite{varian2018artificial,lambrecht2015can,bajari2018impact}, which studied learning without competition, and found that small amounts of additional data do not provide significant improvement in eventual outcomes. We conclude that competition dynamics -- that firms compete as they learn over time -- are pertinent to these anti-trust considerations.

We also investigate how algorithms' performance ``in isolation" (without competition) is predictive of the outcomes under competition. We find that mean reputation -- arguably, the most natural performance measure ``in isolation" -- is sometimes not a good predictor. We suggest a
more refined performance measure, and use it to explain some of the competition outcomes.

We also consider an alternative choice rule with explicit noise/randomness: a small fraction of users choose a firm uniformly at random.%
\footnote{\asedit{Reputation scores already introduce some noise into users' choices. However, the amount of noise due to this channel is typically small, both in our simulations and in practice, because reputation signals average over many datapoints.}} We confirm the theoretical intuition that better algorithms prevail if the expected number of ``random" users is sufficiently large. However, we find that this effect is negligible for some smaller but still ``relevant" parameter values.

\subsection{Discussion}

The present paper is an experimental counterpart to \cite{CompetingBandits-itcs18}, which considered a similar duopoly model and obtained a number of theoretical results with ``asymptotic" flavor. For the sake of analytical tractability, that paper makes a somewhat unrealistic simplification that users do not observe any signals about firms' ongoing performance. Instead, users choose between firms according to the firms' Bayesian-expected rewards. The strength of competition is varied in a different way, using assumptions about (ir)rational user behavior. \asedit{For these reasons, the theorems from \cite{CompetingBandits-itcs18} have no direct bearing on our simulations.} However, their high-level conclusion in is an inverted-U relationship similar to ours.

The present paper provides a more nuanced and ``non-asymptotic" perspective. In essence, we look for substantial effects within relevant time scales. Indeed, we start our investigation by determining what time scales are relevant in the context of our model. The reputation-based choice model accounts for competition in a more direct way, allows to separate reputation vs. data advantage, and makes our model amenable to numerical simulations (unlike the model in \cite{CompetingBandits-itcs18}).

To elucidate the interplay of competition and exploration, our model is stylized in several important  respects, some of which we discuss below. Firms compete only on the quality of service, rather than, say, pricing or the range of products. Various performance signals available to the users, from personal experience to word-of-mouth to consumer reports, are abstracted as persistent ``reputation scores", which further simplified to average performance over a time window.  On the machine learning side, our setup is geared to distinguish between ``simple" vs. ``better" vs. ``smart" bandit algorithms; we are not interested in state-of-art algorithms for very realistic bandit settings.

Even with a stylized model, numerical investigation is quite challenging. An ``atomic experiment" is a competition game between a given pair of bandit algorithms, in a given competition model, on a given instance of a multi-armed bandit problem.%
\footnote{Each such experiment is run many times to reduce variance.}
Accordingly, we have a three-dimensional space of atomic experiments one needs to run and interpret: \{pairs of algorithms\} x \{competition models\} x \{bandit instances\}, and we are looking for findings that are consistent across this entire space. It is essential to keep each of the three dimensions small yet representative. In particular, we need to capture a huge variety of bandit instances with only a few representative examples. Further, one needs succinct and informative summarization of results within one atomic experiment and across multiple experiments (\eg see Table~\ref{sim_table}).

While amenable to simulations, our model appears difficult to analyze. This is for several reasons:
intricate feedback loop from performance to reputations to users to performance;
mean reputation, most connected to our intuition, is sometimes a bad predictor in competition (see Sections~\ref{sec:isolation} and~\ref{sec:revisited});
mathematical tools from regret-minimization would only produce ``asymptotic" results, which do not seem to suffice. Further, we believe that resolving first-order theoretical questions about our model would not add much value to this paper. \asedit{Indeed, we are in the realm of stylized economic models that provide mathematical intuition about the world, and \cite{CompetingBandits-itcs18} already has an elaborate analysis with similar high-level conclusions.}

\subsection{Related work}

\xhdr{Machine learning.} Multi-armed bandits (MAB) is a tractable abstraction for the tradeoff between exploration and exploitation. MAB problems have been studied for many decades, see \cite{Bubeck-survey12,LS19bandit-book} for background on this immense body of work; we only comment on the most related aspects.

We consider MAB with i.i.d. rewards, a well-studied and well-understood MAB model \cite{bandits-ucb1}. We focus on a well-known distinction between ``greedy" (exploitation-only) algorithms, ``naive" algorithms that separate exploration and exploitation, and ``smart" algorithms that combine them. Switching from ``greedy" to ``naive" to ``smart" algorithms involves substantial adoption costs in infrastructure and personnel training \cite{MWT-WhitePaper-2016,DS-arxiv}.

The study of competition vs. exploration has been initiated in \cite{CompetingBandits-itcs18}, as discussed above.
Our setting is also closely related to the ``dueling algorithms" framework \cite{DuelingAlgs-stoc11}, but this framework considers offline / full feedback scenarios whereas we focus on online machine learning problems.

In ``dueling bandits" (e.g., \cite{Yue-dueling-icml09, Yue-dueling12}), an algorithm sets up a ``duel" between a pair of arms in each round, and only learns which arm has ``won". While this setting features a form of competition inside an MAB problem, it is very different from ours.

The interplay between exploration, exploitation and incentives has been studied in other scenarios: incentivizing exploration in a recommendation system,
    \eg \cite{Kremer-JPE14,Frazier-ec14,Che-13,ICexploration-ec15,Bimpikis-exploration-ms17},
dynamic auctions
    (see \cite{DynAuctions-survey10} for background),
online ad auctions, \eg
    \cite{MechMAB-ec09,DevanurK09,NSV08,Transform-ec10-jacm},
human computation
    \cite{RepeatedPA-ec14,Ghosh-itcs13,Krause-www13},
and repeated auctions, \eg
    \cite{Amin-auctions-nips13,Amin-auctions-nips14,Jieming-ec18}.

\xhdr{Economics.}
Our work is related to a longstanding economics literature on competition vs. innovation, \eg \cite{Schumpeter-42,barro2004economic,Aghion-QJE05}. While this literature focuses on R\&D costs of innovation, ``reputational costs" seem new and specific to exploration.

Whether data gives competitive advantage
has been studied theoretically in \cite{varian2018artificial,
  lambrecht2015can} and empirically in
\cite{bajari2018impact}.
While these papers find that small amounts
  of additional data do not provide significant improvement,
 % in training a more predictive model,
  they focus on learning in isolation.
  %, which overlooks certain aspects in a competition.
%\swdelete{These studies consider the effect of additional data in terms of helping towards solving a machine learning problem in isolation and argue that it does not play a large role.}
The first-mover advantage has been well-studied in other settings in economics and marketing, see survey \cite{kerin1992first}.

\asedit{The most common measures of market ``competitiveness" such as the Lerner Index or the Herfindahl-Hirschman Index of a market rely on ex-post observable attributes of a market such as prices or market shares \cite{tirole1988theory}. However, neither is applicable to our setting: in our model, there are no prices, and market shares are endogenous.}

\section{Model and Preliminaries}\label{sec:model}

%\textbf{Overview}
We consider a game involving two firms and $T$ customers (henceforth, \emph{agents}). The game lasts for $T$ rounds. In each round, a new agent arrives, chooses among the two firms, interacts with the chosen firm, and leaves forever.

Each interaction between a firm and an agent proceeds as follows. There is a set $A$ of $K$ actions, henceforth \emph{arms}, same for both firms and all rounds. The firm chooses an arm, and the agent experiences a numerical reward observed by the firm. Each arm corresponds to a different version of the experience that a firm can provide for an agent, and the reward corresponds to the agent's satisfaction level. The other firm does not observe anything about this interaction, not even the fact that this interaction has happened.

From each firm's perspective, the interactions with agents follow the protocol of the multi-armed bandit problem (MAB). We focus on i.i.d. Bernoulli rewards: the  reward of each arm $a$ is drawn from $\{0,1\}$ independently with expectation $\mu(a)$. The mean rewards $\mu(a)$ are the same for all rounds and both firms, but initially unknown.

Before the game starts, each firm commits to an MAB algorithm, and uses this algorithm to choose its actions. Each algorithm receives a ``warm start": additional $T_0$ agents that arrive before the game starts, and interact with the firm as described above. The warm start ensures that each firm has a meaningful reputation when competition starts. Each firm's objective is to maximize its market share: the fraction of users who chose this firm.

In some of our experiments, one firm is the ``incumbent" who enters the market before the other (``late entrant"), and therefore enjoys a \emph{temporary monopoly}. Formally, the incumbent enjoys additional $X$ rounds of the ``warm start". We treat $X$ as an exogenous element of the model, and study the consequences for a fixed $X$.

\xhdr{Agents.}
Firms compete on a single dimension, quality of service, as expressed by agents' rewards. 
\OMIT{\footnote{One can think that there are two quality types - a reward of 1 is a high type and a reward of 0 is a low type}}
Agents are myopic and non-strategic: they would like to choose among the firms so as to maximize their expected reward (i.e. select the firm with the highest quality), without attempting to influence the firms' learning algorithms or rewards of the future users. Agents are not well-informed: they only receive a rough signal about each firm's performance before they choose a firm, and no other information.

Concretely, each of the two firms has a \emph{reputation score}, and each agent's choice is driven by these two numbers. We posit a version of rational behavior: each agent chooses a firm with a maximal reputation score (breaking ties uniformly). The reputation score is simply a sliding window average: an average reward of the last $M$ agents that chose this firm.

%\asedit{This is similar to ratings aggregators on many online platforms such as Yelp, Tripadvisor, etc.}

\xhdr{MAB algorithms.} We consider three classes of algorithms, ranging from more primitive to more sophisticated:

\begin{enumerate}
\item \emph{Greedy algorithms} that strive to take actions with maximal mean reward, based on the current information.

\item \emph{Exploration-separating algorithms} that separate exploration and exploitation. The ``exploitation" choices strives to maximize mean reward in the next round, and the ``exploration" choices do not use the rewards observed so far.
\item \emph{Adaptive exploration}: algorithms that combine exploration and exploitation, and sway the exploration choices towards more promising alternatives.
\end{enumerate}

We are mainly interested in qualitative differences between the three classes. For concreteness, we fix one algorithm from each class. Our pilot experiments indicate that our findings do not change substantially if other algorithms are chosen. For technical reasons, we consider Bayesian versions initialized with a ``fake" prior (\ie not based on actual knowledge). We consider:

\begin{enumerate}
\item a greedy algorithm that chooses an arm with largest posterior mean reward. We call it "Dynamic Greedy" (because the chosen arm may change over time), \DG in short.

\item an exploration-separated algorithm that in each round, \emph{explores} with probability $\eps$: chooses an arm independently and uniformly at random, and with the remaining probability \emph{exploits} according to \DG. We call it ``dynamic epsilon-greedy", \DEG in short.\gaedit{\footnote{Throughout, we fix $\eps = 0.05$. Our pilot experiments showed that different $\eps$ did not qualitatively change the results.}}

\item an adaptive-exploration algorithm called ``Thompson Sampling" (\TS). In each round, this algorithm updates the posterior distribution for the mean reward of each arm $a$, draws an independent sample $s_a$ from this distribution, and chooses an arm with the largest $s_a$.
\end{enumerate}

For ease of comparison, all three algorithms are parameterized with the same fake prior: namely, the mean reward of each arm is drawn independently from a $\Beta(1,1)$ distribution. Recall that Beta priors with 0-1 rewards form a conjugate family, which allows for simple posterior updates.

Both \DEG and \TS are classic and well-understood MAB algorithms, see \cite{Bubeck-survey12,TS-survey-FTML18} for background. It is well-known that \TS is near-optimal in terms of the cumulative rewards, and \DEG is very suboptimal, but still much better than \DG.%
\footnote{Formally, \TS achieves regret
    $\tilde{O}(\sqrt{TK})$ and
    $O(\tfrac{1}{\Delta} \log T)$,
where $\Delta$ is the gap in mean rewards between the best and second-best arms. \DEG has regret $\tilde{\Theta}(T^{2/3} K^{1/3})$ in the worst case. And \DG can have regret as high as $\Omega(T)$. Deeper discussion of these distinctions is not very relevant to this paper.}
In a stylized formula:
    $ \TS \gg \DEG \gg \DG $
as stand-alone MAB algorithms.

\xhdr{MAB instances.}
We consider instances with $K=10$ arms. Since we focus on 0-1 rewards, an instance of the MAB problem is specified by the \emph{\MRV} $(\mu(a):\; a\in A)$. Initially this vector is drawn from some distribution, termed \emph{MAB instance}. We consider three MAB instances:
\begin{enumerate}
\item \emph{Needle-In-Haystack}: one arm (the ``needle") is chosen uniformly at random. This arm has mean reward $.7$, and the remaining ones have mean reward $.5$.

\item \emph{Uniform instance}: the mean reward of each arm is drawn independently and uniformly from $[\nicefrac{1}{4}, \nicefrac{3}{4}]$.
\item \emph{Heavy-Tail instance}: the mean reward of each arm is drawn independently from $\Beta(.6,.6)$ distribution (which is known to have substantial ``tail probabilities").
\end{enumerate}
We argue that these MAB instances are (somewhat) representative. Consider the ``gap" between the best and the second-best arm, an essential parameter in the literature on MAB. The ``gap" is fixed in Needle-in-Haystack, spread over a wide spectrum of values under the Uniform instance, and is spread but  focused on the large values under the Heavy-Tail instance. We also ran smaller experiments with versions of these instances, and achieved similar qualitative results.

\xhdr{Terminology.}
Following a standard game-theoretic terminology, algorithm Alg1 \emph{(weakly) dominates} algorithm Alg2 for a given firm if Alg1 provides a larger (or equal) market share than Alg2 at the end of the game. An algorithm is a (weakly) dominant strategy for the firm if it (weakly) dominates all other algorithms. This is for a particular MAB instance and a particular selection of the game parameters.

\OMIT{\gaedit{Since the algorithms that involve exploration have larger deployment costs, we break any indifference towards algorithms that do not involve exploration (i.e. if \DG and \DEG provide the same market share, the firm will play \DG).}}

\xhdr{Simulation details.}
For each MAB instance we draw $N = 1000$ \MRVs independently from the corresponding distribution. We use this same collection of \MRVs for all experiments with this MAB instance. For each \MRV we draw a table of realized rewards (\emph{realization table}), and use this same table for all experiments on this \MRV. This ensures that differences in algorithm performance are not due to noise in the realizations but due to differences in the algorithms in the different experimental settings.

More specifically, the realization table is a 0-1 matrix $W$ with $K$ columns which correspond to arms, and $T+T_{\max}$ rows, which correspond to rounds. Here $T_{\max}$ is the maximal duration of the ``warm start" in our experiments, \ie the maximal value of $X+T_0$. For each arm $a$, each value $W(\cdot,a)$ is drawn independently from Bernoulli distribution with expectation $\mu(a)$. Then in each experiment, the reward of this arm in round $t$ of the warm start is taken to be $W(t,a)$, and its reward in round $t$ of the game is $W(T_{\max}+t,a)$.

We fix the sliding window size $M = 100$. We found that lower values induced too much random noise in the results, and increasing $M$ further did not make a qualitative difference. Unless otherwise noted, we used $T = 2000$.

The simulations are computationally intensive. An experiment on a particular MAB instance comprised multiple runs of the competition game: $N$ mean reward vectors times $9$ pairs of algorithms times three values for the warm start. We used a parallel implementation over a cluster of 12 2.2 GHz CPU cores, with 8 GB RAM per core. Each experiment took about $10$ hours.

\xhdr{Consistency.}
While we experiment with various MAB instances and parameter settings, we only report on selected, representative experiments in the body of the paper. Additional plots and tables are provided in the appendix. Unless noted otherwise, our findings are based on and consistent with all these experiments.

\section{Performance in Isolation}\label{sec:isolation}

\begin{figure}
\includegraphics[scale=0.35]{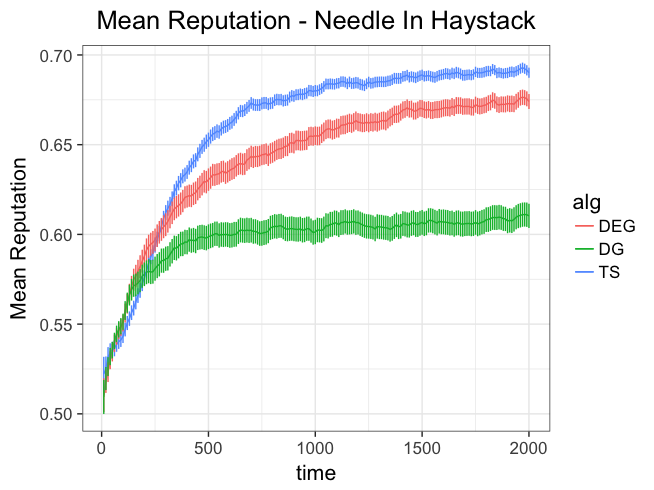}
%\caption*{\tiny{Mean reputation trajectory The plots contain the average reputation over $1000$ runs for a memory size of $100$ where, for a given $t$, we record the reputation of a given algorithm on a given instance and then average this value across all the runs. The shaded area display 95\% confidence intervals.}}
\caption{Mean reputation trajectories for Needle-in-Haystack. The shaded area shows 95\% confidence intervals.}
\label{prelim_means}
\end{figure}

We start with a pilot experiment in which we investigate each algorithm's performance ``in isolation": in a stand-alone MAB problem without competition. We focus on reputation scores generated by each algorithm. We confirm that algorithms' performance is ordered as we'd expect:
    $\TS > \DEG > \DG$
for a sufficiently long time horizon. For each algorithm and each MAB instance, we compute the mean reputation score at each round, averaged over all \MRVs. We plot the \emph{mean reputation trajectory}: how this score evolves over time. Figure \ref{prelim_means} shows such a plot for the Needle-in-Haystack instance; for other MAB instances the plots are similar. We summarize this finding as follows:

\begin{finding}
\textit{The mean reputation trajectories are arranged as predicted by prior work:
    $\TS > \DEG > \DG$ for a sufficiently long time horizon.}
\end{finding}

We also use Figure~\ref{prelim_means} to choose a reasonable time-horizon for the subsequent experiments, as $T=2000$. The idea is, we want $T$ to be large enough so that algorithms performance starts to plateau, but small enough such that algorithms are still learning.

The mean reputation trajectory is probably the most natural way to represent an algorithm's performance on a given MAB instance. However, we found that the outcomes of the competition game are better explained with a different ``performance-in-isolation" statistic that is more directly connected to the game. Consider the  performance of two algorithms, Alg1 and Alg2, ``in isolation" on a particular MAB instance. The \emph{relative reputation} of Alg1 (vs. Alg2) at a given time $t$ is the fraction of \MRVs /realization tables for which Alg1 has a higher reputation score than Alg2. The intuition is that agent's selection in our model depends only on the comparison between the reputation scores.

\begin{figure}[ht]
\includegraphics[scale=0.35]{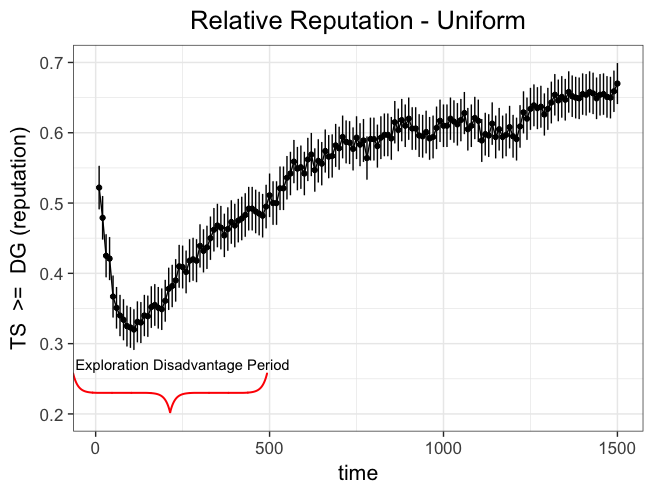}
\includegraphics[scale=0.35]{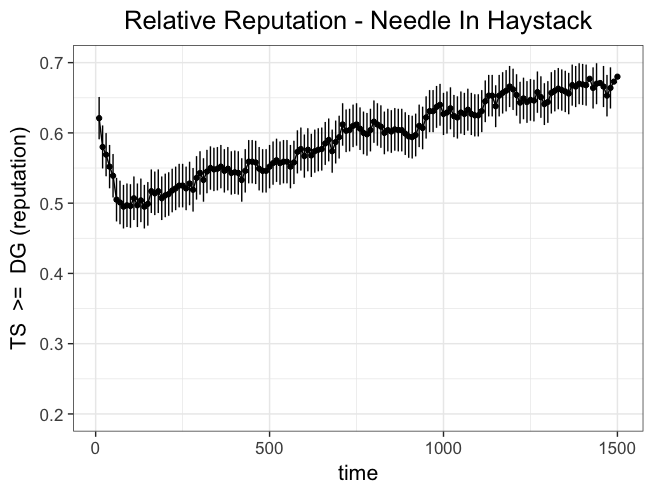}
%\caption*{\tiny{The plots contain the average reputation over $1000$ runs for a memory size of $100$ where, for a given $t$, we record the reputation of both of the algorithms on a given instance and then calculate the proportion of runs where $\TS \geq \DG$. The shaded area display 95\% confidence intervals.}}
\caption{Relative reputation trajectory for $\TS$ vs $\DG$, on Uniform instance (top) and Needle-in-Haystack instance (bottom). Shaded area display 95\% confidence intervals.}
\label{relative_rep_plots}
\end{figure}

This angle allows a more nuanced analysis of reputation costs vs. benefits under competition. Figure \ref{relative_rep_plots} (top) shows the relative reputation trajectory for $\TS$ vs $\DG$ for the Uniform instance. The relative reputation is less than $\tfrac12$ in the early rounds, meaning that $\DG$ has a higher reputation score in a majority of the simulations, and more than $\tfrac12$ later on. The reason is the exploration in \TS leads to worse decisions initially, but allows for better decisions later. The time period when relative reputation vs. \DG dips below $\tfrac12$ can be seen as an explanation for the competitive disadvantage of exploration. Such period also exists for the Heavy-Tail MAB instance. However, it does not exist for the Needle-in-Haystack instance, see Figure \ref{relative_rep_plots} (bottom).%
\footnote{We see two explanations for this: $\TS$ identifies the best arm faster for the Needle-in-Haystack instance, and there are no ``very bad" arms which make exploration very expensive in the short term.}

\begin{finding}\label{find:period}
\textit{Exploration can lead to relative reputation vs. $\DG$ going below $\tfrac12$ for some initial time period. This happens for some MAB instances but not for some others.}
\end{finding}

\begin{definition}
For a particular MAB algorithm, a time period when relative reputation vs. \DG goes below $\tfrac12$ is called {\em exploration disadvantage period}. An MAB instance is called \emph{exploration-disadvantaged} if such period exists.
\end{definition}

\noindent Uniform and Heavy-tail instance are exploration-disadvantaged, but Needle-in-Haystack is not.

\section{Competition vs. Better Algorithms}\label{sec:competition}

% latex table generated in R 3.4.0 by xtable 1.8-2 package
% Thu Aug 16 13:13:01 2018
\footnotesize
\begin{table*}[t]
\centering
\begin{tabular}{|c|c|c|c||c|c|c|}
  \hline
  & \multicolumn{3}{c||}{Heavy-Tail}
  & \multicolumn{3}{c|}{Needle-in-Haystack}\\
  \hline
  & $T_0$ = 20 & $T_0$ = 250 & $T_0$ = 500
  & $T_0$ = 20 & $T_0$ = 250 & $T_0$ = 500 \\
  \hline
\TS vs \DG
  & \makecell{\textbf{0.29} $\pm$0.03\\ \Eeog 55 (0)}
  & \makecell{\textbf{0.72} $\pm$0.02\\ \Eeog 570 (0)}
  & \makecell{\textbf{0.76} $\pm$0.02\\ \Eeog 620 (99)}
  & \makecell{\textbf{0.64} $\pm$0.03\\ \Eeog 200 (27)}
  & \makecell{\textbf{0.6} $\pm$0.03\\ \Eeog 370 (0)}
  & \makecell{\textbf{0.64} $\pm$0.03\\ \Eeog 580 (122)}
  \\
\hline
  $\TS$ vs $\DEG$
  & \makecell{\textbf{0.3} $\pm$0.03\\ \Eeog 37 (0)}
  & \makecell{\textbf{0.88} $\pm$0.01\\ \Eeog 480 (0)}
  & \makecell{\textbf{0.9} $\pm$0.01\\ \Eeog 570 (114)}
  & \makecell{\textbf{0.57} $\pm$0.03\\ \Eeog 150 (14)}
  & \makecell{\textbf{0.52} $\pm$0.03\\ \Eeog 460 (79)}
  & \makecell{\textbf{0.56} $\pm$0.02\\ \Eeog 740 (628)}
  \\
\hline
  $\DG$ vs $\DEG$
  & \makecell{\textbf{0.62} $\pm$0.03\\ \Eeog 410 (7)}
  & \makecell{\textbf{0.6} $\pm$0.02\\ \Eeog 790 (762)}
  & \makecell{\textbf{0.57} $\pm$0.03\\ \Eeog 730 (608)}
  & \makecell{\textbf{0.46} $\pm$0.03\\ \Eeog 340 (129)}
  & \makecell{\textbf{0.42} $\pm$0.02\\ \Eeog 650 (408)}
  & \makecell{\textbf{0.42} $\pm$0.02\\ \Eeog 690 (467)}
  \\
   \hline
\end{tabular}
\normalsize
\caption{{\bf Permanent duopoly}, for Heavy-Tail and Needle-in-Haystack instances. Each cell describes a game between two algorithms, call them Alg1 vs. Alg2, for a particular value of the warm start $T_0$. Line 1 in the cell is the market share of Alg 1: the average (in bold) and the 95\% confidence band.
%For example, the cell in the top left indicates that TS gets on average 64\% of the market when played against DG.
Line 2 specifies the ``effective end of game" (\Eeog): the average and the median (in brackets). The time horizon is $T=2000$.}
\label{sim_table}
\end{table*}

\normalsize
Our main experiments are with the duopoly game defined in Section~\ref{sec:model}. As the ``intensity of competition" varies from permanent monopoly to ``incumbent" to permanent duopoly to ``late entrant", we find a stylized inverted-U relationship as in Figure~\ref{fig:inverted-U}. More formally, we look for equilibria in the duopoly game, where each firm's choices are limited to \DG, \DEG and \TS. We do this for each ``intensity level" and each MAB instance, and look for findings that are consistent across MAB instances. For cleaner results, we break ties towards less advanced algorithms (as they tend to have lower adoption costs \cite{MWT-WhitePaper-2016,DS-arxiv}). Note that \DG is trivially the dominant strategy under permanent monopoly.

%(1) Permanent monopoly - indifferent between all algorithms but break indifference towards DG cause of deployment costs
%(2) Temporary monopoly - TS is the dominant strategy for the incumbent
%(3) Permanent duopoly -
%   (a) Under NIH, TS is dominant
%   (b) Under Heavy Tail, DG is weakly dominant
%   (c) Under Uniform, DG "beats" TS, but DEG "beats" TS by more than DG and DEG vs DG leads to ~50% market share (it's slightly in favor of DG but 50/50 is within the confidence band). If we break indifference towards DG cause of deployment costs, then we have that the unique PSNE is (DG, DG). If we don't add the indifference breaking, then we have four PSNE, (DG, DEG), (DEG, DG), (DG, DG), (DEG, DEG)

\xhdr{Permanent duopoly.}
The basic scenario is when both firms are competing from round $1$. A crucial distinction is whether an MAB instance is exploration-disadvantaged:

\begin{finding}\label{find:duopoly}
\textit{Under permanent duopoly:
\begin{itemize}
\item[(a)] (\DG,\DG) is the unique pure-strategy Nash equilibrium for exploration-disadvantaged MAB instances with a sufficiently small ``warm start".
\item[(b)] This is not necessarily the case for MAB instances that are not exploration-disadvantaged. In particular, \TS is a weakly dominant strategy for Needle-in-Haystack.
\end{itemize}
}
\end{finding}

%\begin{finding}
%\textit{For sufficiently low warm start under the instances that are exploration-disadvantaged, the unique incentivized strategies in the competition game are:\footnote{The uniqueness comes from the fact that we break indifferences towards easier to deploy algorithms}
%\begin{center}
%\textbf{Permanent Monopoly} - $\DG$ \\
%\textbf{Temporary Monopoly} (incumbent)- $\TS$ \\
%\textbf{Permanent Duopoly} - $(\DG, \DG)$ \\
%\end{center}
%The conditions on $\TS$ being dominant under the temporary monopoly are that the incumbent is a temporary monopoly for sufficiently many periods.}
%\end{finding}

%\xhdr{Permanent Monopoly.} Since there is only a single firm in the
%market for the entire period, the firm can take the entire market
%regardless of what algorithm it deploys. \swedit{Since we assume that exploration algorithms ($\DEG$ and $\TS$) would incur deployment cost and firms break indifferences towards easier to deploy algorithms}, the firm would choose to deploy $\DG$.

We investigate the firms' market shares when they choose different algorithms (otherwise, by symmetry both firms get half of the agents). We report the market shares for Heavy-Tail and Needle-in-Haystack instances in Table \ref{sim_table}  (see the first line in each cell), for a range of values of the warm start $T_0$. Table~\ref{tab:duopoly_unif} reports similarly on the Uniform instance. We find that $\DG$ is a weakly dominant strategy for the Heavy-Tail and Uniform instances, as long as $T_0$ is sufficiently small. However, \TS is a weakly dominant strategy for the Needle-in-Haystack instance. We find that for a sufficiently small $T_0$, \DG yields more than half the market against \TS,  but achieves similar market share vs. \DG and \DEG. By our tie-breaking rule, (\DG,\DG) is the only pure-strategy equilibrium.

\begin{table}[h]
\centering
\begin{tabular}{|c|c|c|c|}
 % \hline
 % & \multicolumn{3}{c|}{Uniform} \\
\hline
   & $T_0$ = 20 & $T_0$ = 250 & $T_0$ = 500 \\ \hline
\TS vs \DG
  & \makecell{\textbf{0.46} $\pm$0.03}
    & \makecell{\textbf{0.52} $\pm$0.02}
    & \makecell{\textbf{0.6} $\pm$0.02} \\ \hline
\TS vs \DEG
    & \makecell{\textbf{0.41} $\pm$0.03}
    & \makecell{\textbf{0.51} $\pm$0.02}
    & \makecell{\textbf{0.55} $\pm$0.02} \\ \hline
\DG vs \DEG
    & \makecell{\textbf{0.51} $\pm$0.03}
    & \makecell{\textbf{0.48} $\pm$0.02}
    & \makecell{\textbf{0.45} $\pm$0.02} \\\hline
\end{tabular}
\caption{{\bf Permanent duopoly}, for the Uniform MAB instance. Semantics are the same as in Table \ref{sim_table}.}
\label{tab:duopoly_unif}
\end{table}

\OMIT{This establishes our claim that $\DG$ is the incentivized algorithm.\footnote{We defer the table for uniform instances to the appendix. Summarizing the results, we see that, for low warm start, \DG yields more than half the market against \TS but is indifferent between \DG and \DEG. Since we break indifference towards easier to deploy algorithms, we also find that \DG is the incentivized algorithm in equilibrium}}

We attribute the prevalence of \DG on exploration-disadvantaged MAB instances to its prevalence on the initial ``exploration disadvantage period", as described in Section~\ref{sec:isolation}. Increasing the warm start length $T_0$ makes this period shorter: indeed, considering relative reputation trajectory in Figure~\ref{relative_rep_plots} (top), increasing $T_0$ effectively shifts the starting time point to the right. This is why it helps \DG if $T_0$ is small.

\OMIT{
The reason we see that $\DG$ is incentivized under the Heavy Tail and Uniform instances but that $\TS$ is weakly dominant under Needle In Haystack is precisely due to the fact that the purposeful exploration engaged by $\TS$ in the early rounds puts it in a disadvantage in the former case but not in the latter. Looking at the relative reputation plots in Figure \ref{relative_rep_plots}, we can interpret fixing a warm start $T_0$ as fixing the starting point on the relative reputation plots. The proportion of first rounds in the competition game that will go to a firm playing alg $A$ over a firm playing alg $B$ will correspond to the relative reputation proportion at time $T_0$. As a result, we see that for warm start $T_0 = 20$ on the exploration-disadvantaged instances, $\DG$ will win the first agent in a larger proportion of the simulations than $\TS$. However, on the Needle In Haystack instances, we see that $\TS$ wins a larger proportion of the first agents compared to $\DG$.
}

\xhdr{Temporary Monopoly.}
We turn our attention to the temporary monopoly scenario. Recall that the incumbent firm enters the market and serves as a monopolist until the entrant firm enters at round $X$. We make $X$ large enough, but still much smaller than the time horizon $T$. We find that the incumbent is incentivized to choose \TS, in a strong sense:

\begin{finding}\label{find:temp-monopoly}
\textit{Under temporary monopoly, \TS is the dominant strategy for the incumbent. This holds across all MAB instances, if $X$ is large enough.
}
\end{finding}

The simulation results for the Heavy-Tail MAB instance are reported in Table~\ref{tab:ht-incum}, for a particular $X=200$. We see that \TS is a dominant strategy for the incumbent. Similar tables for the other MAB instances and other values of $X$ are reported in the supplement, with the same conclusion.

\begin{table}[H]
\centering
\begin{tabular}{|c|c|c|c|}
\hline
   & $\TS$  & $\DEG$  & $\DG$ \\ \hline
$\TS$
    & \makecell{\textbf{0.003}$\pm$0.003}
    & \makecell{\textbf{0.083}$\pm$0.02}
    & \makecell{\textbf{0.17}$\pm$0.02} \\\hline
$\DEG$
    & \makecell{\textbf{0.045}$\pm$0.01}
    & \makecell{\textbf{0.25}$\pm$0.02}
    & \makecell{\textbf{0.23}$\pm$0.02} \\\hline
$\DG$
    & \makecell{\textbf{0.12}$\pm$0.02}
    & \makecell{\textbf{0.36}$\pm$0.03}
    & \makecell{\textbf{0.3}$\pm$0.02} \\\hline
\end{tabular}
\caption{{\bf Temporary monopoly}, with $X=200$ (and $T_0=20$), for the Heavy-Tail MAB instance. Each cell describes the duopoly game between the entrant's algorithm (the row) and the incumbent's algorithm (the column). The cell specifies the entrant's market share (fraction of rounds in which it was chosen) for the rounds in which he was present. We give the average (in bold) and the 95\% confidence interval. NB: smaller average is better for the incumbent.}
\label{tab:ht-incum}
\end{table}
\OMIT{
\begin{table}[h]
\centering
\caption{Heavy Tail}
\begin{tabular}{|c|c|c|c|}
\hline
   & $\TS$  & $\DEG$  & $\DG$ \\ \hline
$\TS$
    & \makecell{\textbf{0.50, 0.50}}
    & \makecell{\textbf{0.3, 0.7}}
    & \makecell{\textbf{0.29, 0.71}} \\\hline
$\DEG$
    & \makecell{\textbf{0.7, 0.3}}
    & \makecell{\textbf{0.50, 0.50}}
    & \makecell{\textbf{0.38, 0.62}} \\\hline
$\DG$
    & \makecell{\textbf{0.71, 0.29}}
    & \makecell{\textbf{0.62, 0.38}}
    & \makecell{\textbf{0.50, 0.50}} \\\hline
\end{tabular}
\end{table}

\begin{table}[h]
\centering
\caption{Needle In Haystack}
\begin{tabular}{|c|c|c|c|}
\hline
   & $\TS$  & $\DEG$  & $\DG$ \\ \hline
$\TS$
    & \makecell{\textbf{0.50, 0.50}}
    & \makecell{\textbf{0.57, 0.43}}
    & \makecell{\textbf{0.64, 0.36}} \\\hline
$\DEG$
    & \makecell{\textbf{0.43, 0.57}}
    & \makecell{\textbf{0.50, 0.50}}
    & \makecell{\textbf{0.54, 0.46}} \\\hline
$\DG$
    & \makecell{\textbf{0.36, 0.64}}
    & \makecell{\textbf{0.43, 0.57}}
    & \makecell{\textbf{0.50, 0.50}} \\\hline
\end{tabular}
\end{table}
}

\DG is a weakly dominant strategy for the entrant, for Heavy-Tail instance in Table~\ref{tab:ht-incum} and the Uniform instance, but not for the Needle-in-Haystack instance. We attribute this finding to exploration-disadvantaged property of these two MAB instance, for the same reasons as discussed above.

\begin{finding}\label{find:temp-monopoly-entrant}
\textit{Under temporary monopoly, \DG is a weakly dominant strategy for the entrant for exploration-disadvantaged MAB instances.
}
\end{finding}

\xhdr{Inverted-U relationship.}
We interpret our findings through the lens of the inverted-U relationship between the ``intensity of competition" and the ``quality of technology". The lowest level of competition is monopoly, when \DG wins out for the trivial reason of tie-breaking. The highest levels are permanent duopoly and ``late entrant". We see that \DG is incentivized for exploration-disadvantaged MAB instances. In fact, incentives for \DG get stronger when the model transitions from permanent duopoly to ``late entrant".%
\footnote{For the Heavy-Tail instance, \DG goes from a weakly dominant startegy to a strictly dominant one. For the Uniform instance, \DG goes from a Nash equilibrium strategy to a weakly dominant one.}
Finally, the middle level of competition, ``incumbent" in the temporary monopoly creates strong incentives for \TS. In stylized form, this relationship is captured in Figure~\ref{fig:inverted-U}.

% Transition Duopoly -> LateStart creates stronger incentives for DG.
% In Heavy Tail, DG is strictly dominant (before it was only weakly).
% In Uniform, DG is weakly dominant (before it was only PSNE).

Our intuition for why incumbency creates more incentives for exploration is as follows. During the temporary monopoly period, reputation costs of exploration vanish. Instead, the firm wants to improve its performance as much as possible by the time competition starts. Essentially, the firm only faces a classical explore-exploit tradeoff, and is incentivized to choose algorithms that are best at optimizing this tradeoff.

\xhdr{Death spiral effect.}
Further, we investigate the ``death spiral" effect mentioned in the Introduction. Restated in terms of our model, the effect is that one firm attracts new customers at a lower rate than the other, and falls behind in terms of performance because the other firm has more customers to learn from, and this gets worse over time until (almost) all new customers go to the other firm. With this intuition in mind, we define  \textit{effective end of game} (\Eeog) for a particular \MRV and realization table, as the last round $t$ such that the agents at this and previous round choose different firms. Indeed, the game, effectively, ends after this round. We interpret low \Eeog as a strong evidence of the ``death spiral" effect. Focusing on the permanent duopoly scenario, we specify the \Eeog values in Table \ref{sim_table} (the second line of each cell). We find that the \Eeog values are indeed small:

\begin{finding}
\textit{
Under permanent duopoly, \Eeog values tend to be much smaller than the time horizon $T$.
}
\end{finding}

We also see that the \Eeog values tend to increase as the warm start $T_0$ increases. We conjecture this is because larger $T_0$ tends to be more beneficial for a better algorithm (as it tends to follow a better learning curve). Indeed, we know that the ``effective end of game" in this scenario typically occurs when a better algorithm loses, and helping it delays the loss.

\xhdr{Welfare implications.}
We study the effects of competition on consumer welfare: the total reward collected by the users over time. Rather than welfare directly, we find it more lucid to consider
\emph{market regret}:
\[ \textstyle T\, \max_a \mu(a) - \sum_{t\in [T]} \mu(a_t), \]
where $a_t$ is the arm chosen by agent $t$. This is a standard performance measure in the literature on multi-armed bandits. Note that smaller regret means higher welfare.

\begin{figure}
\includegraphics[scale=0.3]{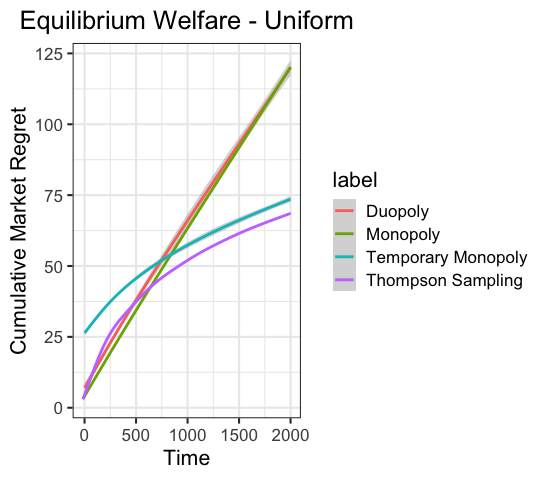}
\includegraphics[scale=0.3]{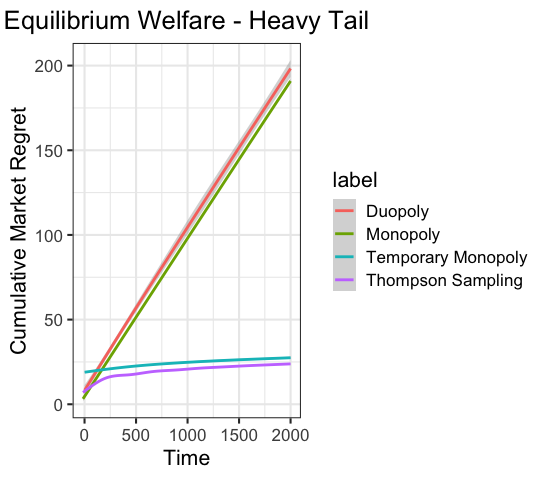}
\caption{Smoothed welfare plots resulting from equilibrium strategies in the different market structures. Note that welfare at $t = 0$ incorporates the regret incurred during the incumbent and warm start periods.}
\label{eq_regret}
\end{figure}

We assume that both firms play their respective equilibrium strategies for the corresponding competition level. As discussed previously, these are:
\begin{itemize}
\item \DG in the monopoly,
\item \DG for both firms in duopoly (Finding \ref{find:duopoly}),
\item \TS for the incumbent (Finding \ref{find:temp-monopoly}) and \DG for the entrant in temporary monopoly (Finding \ref{find:temp-monopoly-entrant}).
\end{itemize}

Figure \ref{eq_regret} displays the market regret (averaged
  over multiple runs) under different levels of competition.
Consumers are \textit{better off} in the temporary monopoly case than in
the duopoly case. Recall that under temporary monopoly, the incumbent is incentivized to play \TS. Moreover, we find that the welfare is close to that of having a single firm for all agents and running \TS. We also observe that monopoly and duopoly achieve similar welfare.
%with monopoly being marginally better than duopoly.

\begin{finding}\label{find:welfare}
\textit{In equilibrium, consumer welfare is (a) highest under temporary monopoly, (b) similar for monopoly and duopoly.
%monopoly being marginally better leading to marginally better welfare than duopoly.
}
\end{finding}

Finding~\ref{find:welfare}(b) is interesting because, in equilibrium, both firms play \DG in both settings, and one might conjecture that the welfare should increase with the number of firms playing \DG. Indeed, one run of \DG may get stuck on a bad arm. However, two firms independently playing \DG are less likely to get stuck simultaneously. If one firm gets stuck and the other does not, then the latter should attract most agents, leading to improved welfare.
\begin{figure}
\includegraphics[scale=0.3]{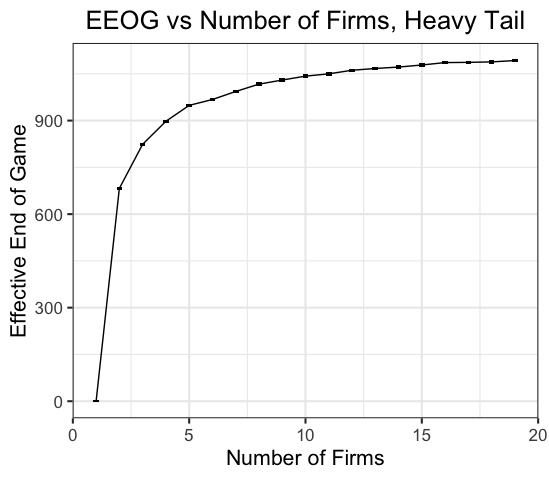}
\includegraphics[scale=0.3]{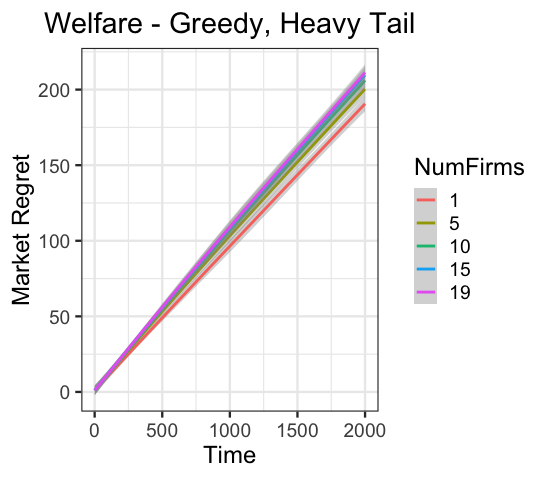}\\
\caption{Average welfare and \Eeog as we increase \#firms playing \DG}
\label{many_firm_welfare}
\end{figure}

To study this phenomenon further, we go beyond the duopoly setting to more than two firms playing \DG (and starting at the same time). Figure \ref{many_firm_welfare} reports the average welfare
across these simulations. Welfare not only does not get better, \textit{but is weakly worse} as we increase the number of firms.

\begin{finding}
\textit{When all firms deploy \DG, and start at the same time, welfare is weakly decreasing as the number of firms \gaedit{increases}}
\end{finding}

We track the average \Eeog in each of the
simulations and notice that it \textit{increases} with the number of firms.
This observation also runs counter of the intuition that with more firms running \DG, one of them is more likely to ``get lucky" and take over the market (which would cause \Eeog to \emph{decrease} with the number of firms).

\OMIT{As we increase the number of firms, a single
individual firm gets fewer consumers for learning as consumers switch
between firms more often. Additionally, bad luck with rewards or arm
selection is punished even more harshly as we increase the number of
firms. This further strengthens the intuition from our previous
result, which is that learning may be hard under competition since
exploration and mistakes are punished harshly but are necessary for
learning.}

\section{Data as a Barrier to Entry}\label{sec:barriers}

\OMIT{
\gaedit{Even though welfare is highest in the temporary monopoly case there are other reasons one may worry about a single firm having a substantial market share. In this section we explore what drives the substantial market share of the temporary monopolist and discuss how this relates to thinking about data as a barrier to entry.}\swcomment{maybe articulate the ``other'' reasons?}
}

%We explore what factors drive the large market share for the incumbent in the temporary monopoly model.

Under temporary monopoly, the incumbent can explore without incurring immediate reputational costs, and build up a high reputation before the entrant appears. Thus, the early entry gives the incumbent both a \textit{data} advantage and a \textit{reputational} advantage over the entrant. We explore which of the two factors is more significant.  Our findings provide a quantitative insight into the role of the classic ``first mover advantage" phenomenon in the digital economy.

%Second, our findings provide a viewpoint in terms of thinking about the role that data can play as a competitive advantage.

For a more succinct terminology, recall that the incumbent enjoys an extended warm start of $X+T_0$ rounds. Call the first $X$ of these rounds the \emph{monopoly period} (and the rest is the proper ``warm start"). The rounds when both firms are competing for customers are called \emph{competition period.}

We run two additional experiments to isolate the effects of the two
advantages mentioned above. The \emph{data-advantage experiment} focuses on the data advantage by, essentially, erasing the reputation advantage. Namely, the data from the monopoly period is not used in the computation of the incumbent's reputation score. Likewise, the \emph{reputation-advantage experiment} erases the data advantage and focuses on the reputation advantage: namely, the incumbent's algorithm `forgets' the data gathered during the monopoly period.

We find that either data or reputational advantage alone gives a substantial boost to the incumbent, compared to permanent duopoly. The results for the Heavy-Tail instance are presented in Table~\ref{barrier_exp}, in the same structure as Table~\ref{tab:ht-incum}. For the other two instances, the results are qualitatively similar.

\begin{table*}[h]
\centering
\begin{tabular}{|c|c|c|c||c|c|c|}
\hline
  & \multicolumn{3}{c||}{Reputation advantage}
  & \multicolumn{3}{c|}{Data advantage}\\
\hline
& $\TS$  & $\DEG$  & $\DG$
& $\TS$  & $\DEG$  & $\DG$
\\\hline
$\TS$
    & \makecell{\textbf{0.021}$\pm$0.009}
    & \makecell{\textbf{0.16}$\pm$0.02}
    & \makecell{\textbf{0.21} $\pm$0.02}
    & \makecell{\textbf{0.0096}$\pm$0.006}
    & \makecell{\textbf{0.11}$\pm$0.02}
    & \makecell{\textbf{0.18}$\pm$0.02}
    \\ \hline
$\DEG$
    & \makecell{\textbf{0.26}$\pm$0.03}
    & \makecell{\textbf{0.3}$\pm$0.02}
    & \makecell{\textbf{0.26}$\pm$0.02}
    & \makecell{\textbf{0.073}$\pm$0.01}
    & \makecell{\textbf{0.29}$\pm$0.02}
    & \makecell{\textbf{0.25}$\pm$0.02}
    \\ \hline
$\DG$
    & \makecell{\textbf{0.34}$\pm$0.03}
    & \makecell{\textbf{0.4}$\pm$0.03}
    & \makecell{\textbf{0.33}$\pm$0.02}
    & \makecell{\textbf{0.15}$\pm$0.02}
    & \makecell{\textbf{0.39}$\pm$0.03}
    & \makecell{\textbf{0.33}$\pm$0.02}
    \\\hline
\end{tabular}
\caption{Data advantage vs. reputation advantage experiment, on Heavy-Tail MAB instance. Each cell describes the duopoly game between the entrant's algorithm (the {\bf row}) and the incumbent's algorithm (the {\bf column}). The cell specifies the entrant's market share for the rounds in which hit was present: the average (in bold) and the 95\% confidence interval. NB: smaller average is better for the incumbent.}
\label{barrier_exp}
\end{table*}

We can quantitatively define the data (resp., reputation) advantage as the incumbent's market share in the competition period in the data-advantage (resp., reputation advantage) experiment, minus the said share under permanent duopoly, for the same pair of algorithms and the same problem instance. In this language, our findings are as follows.

\begin{finding}\label{barrier-find}
\textit{\\
(a) Data advantage and reputation advantage alone are substantially large, across all algorithms and all MAB instances. \\(b) The data advantage is larger than the reputation advantage when the incumbent chooses \TS. \\(c) The two advantages are similar in magnitude when the incumbent chooses \DEG or \DG.
}
\end{finding}

Our intuition for Finding~\ref{barrier-find}(b) is as follows. Suppose the incumbent switches from \DG to \TS. This switch allows the incumbent to explore actions more efficiently -- collect better data in the same number of rounds -- and therefore should benefit the data advantage. However, the same switch increases the reputation cost of exploration in the short run, which could weaken the reputation advantage.

\section{Performance in Isolation, Revisited}\label{sec:revisited}

We saw in Section~\ref{sec:competition} that mean reputation trajectories do not suffice to explain the outcomes under competition. Let us provide more evidence and intuition for this.

Mean reputation trajectories are so natural that one is tempted to conjecture that they determine the outcomes under competition. More specifically:
\begin{conjecture}\label{conj:mean-trajectories}
If one algorithm's mean reputation trajectory lies above another, perhaps after some initial time interval (\eg as in Figure~\ref{prelim_means}), then the first algorithm prevails under competition, for a sufficiently large warm start $T_0$.
\end{conjecture}

However, we find a more nuanced picture. For example, in Figure \ref{sim_table} we see that $\DG$ attains a larger market share than $\DEG$ even for large warm starts. We find that this also holds for $K = 3$ arms and longer time horizons, see the supplement for more details. We conclude:

\begin{finding}
\textit{
Conjecture~\ref{conj:mean-trajectories} is false: mean reputation trajectories do not suffice to explain the outcomes under competition.}
\end{finding}

\begin{figure}[ht]
\includegraphics[scale=0.35]{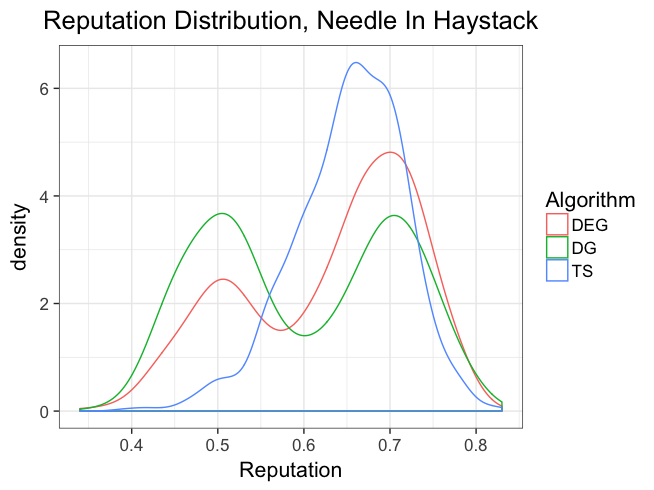}
\caption{Reputation scores for Needle-in-Haystack at $t=500$ (smoothed using a kernel density estimate)}
\label{rep_dist_nih}
%\caption*{\tiny{The plots contain a kernel density estimate of the reputation distribution at $t = 500$}}
\end{figure}

To see what could go wrong with Conjecture~\ref{conj:mean-trajectories}, consider how an algorithm's reputation score is distributed at a particular time. That is, consider the empirical distribution of this score over different \MRVs.%
\footnote{Recall that each \MRV in our experimental setup comes with one specific realization table.} For concreteness, consider the Needle-in-Haystack instance at time $t=500$, plotted in Figure \ref{rep_dist_nih}. (The other MAB instances lead to a similar intuition.)

We see that the ``naive" algorithms $\DG$ and $\DEG$ have a bi-modal reputation distribution, whereas $\TS$ does not. The reason is that for this MAB instance, $\DG$ either finds the best arm and sticks to it, or gets stuck on the bad arms. In the former case \DG does slightly better than $\TS$, and in the latter case it does substantially worse. However, the mean reputation trajectory may fail to capture this complexity since it simply takes average over different \MRVs. This may be inadequate for explaining the outcome of the duopoly game, given that the latter is determined by a simple comparison between the firm's reputation scores.

\begin{figure}[ht]
\includegraphics[scale=0.35]{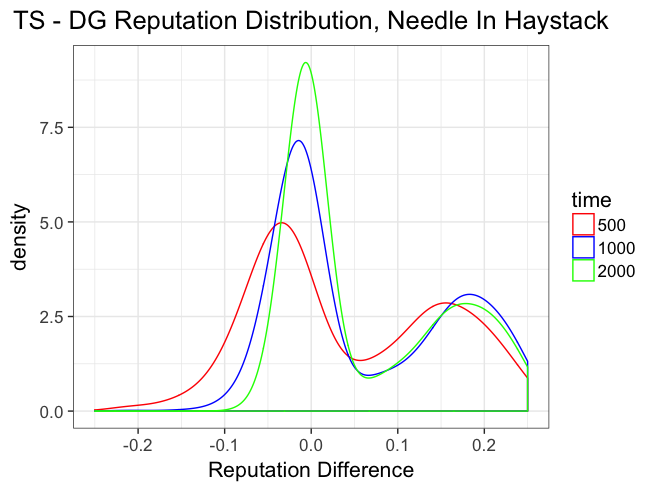}
\caption{Reputation difference $\TS-\DG$ for  Needle-in-Haystack
% at $t = 500, 1000, 2000$ AS: this is redundant, and it saves a line :).
(smoothed using a kernel density estimate)}
\label{ts_dg_rep_diff_nih}
%\caption*{\tiny{The plots contain a kernel density estimate of the difference in reputation between $\TS$ and $\DG$ across $t$}}
\end{figure}

To further this intuition, consider the difference in reputation scores (\emph{reputation difference}) between \TS and \DG on a particular \MRV. Let's plot the empirical distribution of the reputation difference (over the \MRVs) at a particular time point. Figure~\ref{ts_dg_rep_diff_nih} shows such plots for several time points. We observe that the distribution is skewed to the right, precisely due to the fact that $\DG$ either does slightly better than $\TS$ or does substantially worse. Therefore, the mean is not a good measure of the central tendency, or typical value, of this distribution.

\OMIT{In particular, the mean reputation difference tends to overstate how well an advanced exploration algorithm (that will eventually find the best arm) should do in competition compared to a naive algorithm that may never find the best arm.}

\section{Non-deterministic Choice Models}\label{sec:non_greedy}

Let us consider an extension in which the agents' choice is no longer deterministic. Recall that in our main model agents deterministically choose the firm with the higher reputation score; call this choice rule \emph{HardMax} (\HM). Now, we introduce some randomness: each agent selects between the firms uniformly with probability $\eps\in (0,1)$, and takes the firm with the higher reputation score with the remaining probability; call this choice rule \emph{HardMax with randomness} (\HMR).

One can view \HMR as a version of ``warm start", where a firm receives some customers without competition, but these customers are dispersed throughout the game. The expected duration of this ``dispersed warm start" is $\eps T$. If this quantity is large enough, we expect better algorithms to reach their long-term performance and prevail in competition. We confirm this intuition; we also find that this effect is negligible for smaller (but still relevant) values of $\eps$ or $T$.

\OMIT{In Section \ref{sec:competition} we show that in the permanent duopoly
case exploration can lead to a death spiral which eventually starves
the firm of agents. However, giving one firm a small head start or
enough free agents via the warm start incentivizes it to play \TS
since it could recover the reputation costs it incurred from early
exploration. An interpretation of \HMR~ is that instead of
concentrating the free agents as arriving in the beginning of the game
instead they are dispersed throughout the game.}

\begin{finding}\label{find:non_greedy_choice}
\textit{\TS is weakly dominant under the \HMR~ choice rule, if and only if $\eps T$ is sufficiently large. \HMR leads to lower variance in market shares, compared to \HM.}
\end{finding}

\footnotesize
\begin{table*}[t]
\centering
\begin{tabular}{|c|c|c|c||c|c|c|}
  \hline
  & \multicolumn{3}{c||}{Heavy-Tail (\HMR with $\eps=.1$)}
  & \multicolumn{3}{c|}{Heavy-Tail (\HM)}\\
  \hline
  & \TS vs \DG & \TS vs \DEG  & \DG vs \DEG
 & \TS vs \DG & \TS vs \DEG  & \DG vs \DEG  \\
  \hline
$T = 2000$
 & \makecell{ \textbf{0.43} $\pm$ 0.02 \\Var: 0.15 }
  & \makecell{ \textbf{0.44} $\pm$ 0.02 \\Var: 0.15 }
  & \makecell{ \textbf{0.6} $\pm$ 0.02 \\Var: 0.1 }
 &  \makecell{ \textbf{0.29} $\pm$ 0.03 \\Var: 0.2 }
  & \makecell{ \textbf{0.28} $\pm$ 0.03 \\Var: 0.19 }
  & \makecell{ \textbf{0.63} $\pm$ 0.03 \\Var: 0.18 }
    \\
\hline
  $T= 5000$
   & \makecell{ \textbf{0.66} $\pm$ 0.01 \\Var: 0.056 }
  & \makecell{ \textbf{0.59} $\pm$ 0.02 \\Var: 0.092 }
  & \makecell{ \textbf{0.56} $\pm$ 0.02 \\Var: 0.098 }
 & \makecell{ \textbf{0.29} $\pm$ 0.03 \\Var: 0.2 }
 & \makecell{ \textbf{0.29} $\pm$ 0.03 \\Var: 0.2 }
 & \makecell{ \textbf{0.62} $\pm$ 0.03 \\Var: 0.19 }
 \\
  \hline
  $T = 10000$
  & \makecell{ \textbf{0.76} $\pm$ 0.01 \\Var: 0.026 }
 & \makecell{ \textbf{0.67} $\pm$ 0.02 \\Var: 0.067 }
 & \makecell{ \textbf{0.52} $\pm$ 0.02 \\Var: 0.11 }
  & \makecell{ \textbf{0.3} $\pm$ 0.03 \\Var: 0.21 }
  & \makecell{ \textbf{0.3} $\pm$ 0.03 \\Var: 0.2 }
  & \makecell{ \textbf{0.6} $\pm$ 0.03 \\Var: 0.2 }
  \\
   \hline
\end{tabular}
\normalsize
\caption{\HM and \HMR choice models on the Heavy-Tail MAB instance. Each cell describes the market shares in a game between two algorithms, call them Alg1 vs. Alg2, at a particular value of $t$. Line 1 in the cell is the market share of Alg 1: the average (in bold) and the 95\% confidence band.
%For example, the cell in the top left indicates that TS gets on average 64\% of the market when played against DG.
Line 2 specifies the variance of the market shares across the simulations. The results reported here are with $T_0 = 20$.}
\label{tab:non_greedy_table}
\end{table*}

\normalsize

\footnotesize
\begin{table*}[t]
\centering
\begin{tabular}{|c|c|c|c||c|c|c|}
  \hline
  & \multicolumn{3}{c||}{Uniform (\HMR  with $\eps=.1$)}
  & \multicolumn{3}{c|}{Needle-In-Haystack (\HMR  with $\eps=.1$)}\\
  \hline
  & \TS vs \DG & \TS vs \DEG  & \DG vs \DEG
 & \TS vs \DG & \TS vs \DEG  & \DG vs \DEG  \\
 \hline
$T = 2000$
 & \makecell{ \textbf{0.42} $\pm$ 0.02 \\Var: 0.13 }
 & \makecell{ \textbf{0.45} $\pm$ 0.02 \\Var: 0.13 }
 & \makecell{ \textbf{0.49} $\pm$ 0.02 \\Var: 0.093 }
  & \makecell{  \textbf{0.55} $\pm$ 0.02 \\Var: 0.15 }
  & \makecell{  \textbf{0.61} $\pm$ 0.02 \\Var: 0.13 }
  & \makecell{  \textbf{0.46} $\pm$ 0.02 \\Var: 0.12 }
    \\
\hline
  $T= 5000$
 & \makecell{ \textbf{0.48} $\pm$ 0.02 \\Var: 0.089 }
 & \makecell{ \textbf{0.53} $\pm$ 0.02 \\Var: 0.098 }
 & \makecell{ \textbf{0.46} $\pm$ 0.02 \\Var: 0.072 }
 & \makecell{  \textbf{0.56} $\pm$ 0.02 \\Var: 0.13 }
 & \makecell{  \textbf{0.63} $\pm$ 0.02 \\Var: 0.12 }
 & \makecell{  \textbf{0.43} $\pm$ 0.02 \\Var: 0.11 }
 \\
  \hline
  $T = 10000$
& \makecell{ \textbf{0.54} $\pm$ 0.01 \\Var: 0.055 }
& \makecell{  \textbf{0.6} $\pm$ 0.02 \\Var: 0.073 }
& \makecell{  \textbf{0.44} $\pm$ 0.02 \\Var: 0.064 }
  & \makecell{ \textbf{0.58} $\pm$ 0.02 \\Var: 0.083 }
  & \makecell{ \textbf{0.65} $\pm$ 0.02 \\Var: 0.096 }
  & \makecell{ \textbf{0.4} $\pm$ 0.02 \\Var: 0.1 }
  \\
   \hline
\end{tabular}
\caption{\HMR choice model for Uniform and Needle-In-Haystack MAB instances.}
%Same semantics as in Table \ref{tab:non_greedy_table}.
\label{tab:additional_results}
\end{table*}
\normalsize

Table \ref{tab:non_greedy_table} shows the average market shares under
the \HM~ vs \HMR~ choice rule. In contrast to the \HM~ model,
  \TS becomes weakly dominant under the \HMR model, as $T$ gets
  sufficiently large. These findings hold across all problem
instances, see Table \ref{tab:additional_results} (with the same semantics as in Table \ref{tab:non_greedy_table}).

\OMIT{\footnote{The results here are pulled using different \MRV
  and realizations than the results pulled previously (due to the
  larger $T$). However, they are drawn from the same prior instances
  and so qualitatively are the same but the quantitative results are
  not directly comparable to those from the previous
  section.\swcomment{not sure we need this}}}

\normalsize
\OMIT{The intuition for why \TS
should become the dominant strategy eventually is simple. The
consistent stream of random agents ensures that each principal is
chosen at least $\Omega(\eps t)$ times at every time step $t$. As
a result, each algorithm should eventually converge to its asymptotic
performance in isolation.}

\OMIT{Finding \ref{find:non_greedy_choice} implies that we can re-interpret the inverted-U findings from before in terms of the number of agents that a firm receives without having to worry about incentives. In the extreme when the firm gets all agents for free as in the monopoly case then it is incentivized to play \DG. When it only gets some of the agents for free, either via a large warm start, a temporary monopoly, or non-deterministic choice, then \TS is incentivized. However, if the number of free agents gets small enough then \DG is incentivized as in the permanent duopoly analysis from before.\swcomment{I don't understand this paragraph; I think it's too vague. Perhaps remove?}}

However, it takes a significant amount of randomness and a relatively large time horizon for this effect to take place. Even with $T = 10000$ and $\eps = 0.1$ we see that \DEG still outperforms \DG on the Heavy-Tail MAB instance as well as that \TS only starts to become weakly dominant at $T = 10000$ for the Uniform MAB instance.

\OMIT{Table \ref{tab:non_greedy_table} also shows that another difference between the two choice rules is that \HMR leads to lower variance in market shares across simulations compared to \HM.}

\section{Conclusion}\label{sec:conclusion}

We consider a stylized duopoly setting where firms simultaneously learn from and compete for users. We showed that competition may not always induce firms to commit to better exploration algorithms, resulting in welfare losses for consumers. The primary reason is that exploration may have short-term reputational consequences that lead to more naive algorithms winning in a long-term competition. Allowing one firm to have a head start, a.k.a. the first-mover advantage, incentivizes the first-mover to deploy ``better" algorithms, which in turn leads to better welfare for consumers. Finally, we isolate the component of the first-mover advantage that is due to having more initial data, and find that even a small amount of this ``data advantage" leads to substantial long-term market power.
\newpage
\bibliographystyle{acm}
\bibliography{refs,bib-abbrv-short,bib-bandits,bib-AGT,bib-slivkins}

\
\begin{appendices}

We provide plots and tables for our experiments, which were omitted from the main text due to page constraints. In all cases, the plots and tables here are in line with those in the main text, and lead to similar qualitative conclusions.

\section{Plots for ``Performance In Isolation"}

We present additional plots for Section \ref{sec:isolation}. First, we provide mean reputation trajectories for Uniform and Heavy-Tail MAB instances. Second, we provide trajectories for instantaneous mean rewards, for all three MAB instances.%
\footnote{These trajectories are smoothed via a non-parametric regression.
More concretely, we use this option in $\texttt{ggplot}$:
\url{https://ggplot2.tidyverse.org/reference/geom_smooth.html}.}
In all plots, the shaded area represents 95\% confidence interval.

\begin{center}
\includegraphics[scale=0.35]{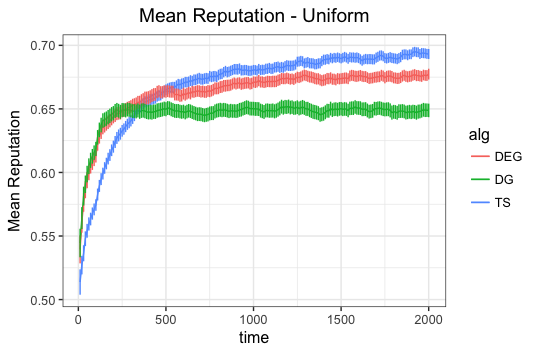}
\includegraphics[scale=0.35]{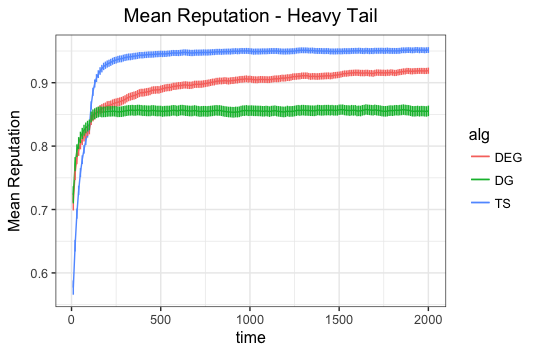}
\end{center}
\begin{center}
\includegraphics[scale=0.35]{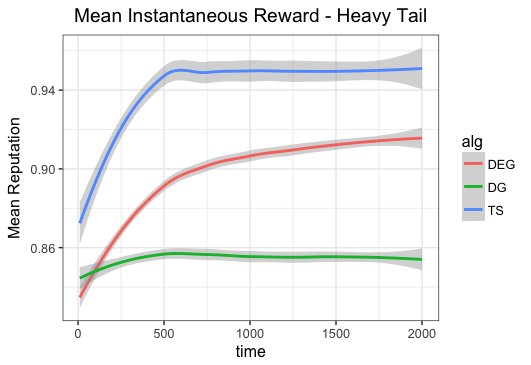}
\includegraphics[scale=0.35]{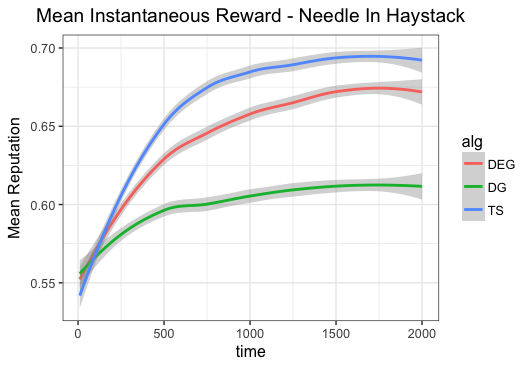}
\end{center}
\begin{center}
\includegraphics[scale=0.35]{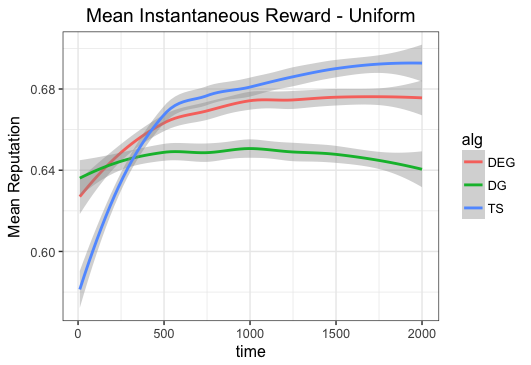}
\end{center}

\section{Temporary Monopoly}

We present additional experiments on temporary monopoly from Section \ref{sec:competition}, across various MAB instances and various values of the incumbent advantage parameter $X$.

Each experiment is presented as a table with the same semantics as in the main text. Namely, each cell in the table describes the duopoly game between the entrant's algorithm (the row) and the incumbent's algorithm (the column). The cell specifies the entrant's market share (fraction of rounds in which it was chosen) for the rounds in which he was present. We give the average (in bold) and the 95\% confidence interval. NB: smaller average is better for the incumbent.

\OMIT{These results confirm the claim in the text that, for sufficiently large $X$, $\TS$ is preferred over all other algorithms for the incumbent. However, it also shows that, for smaller values of $X$ it is not necessarily the case that $\TS$ is the preferred algorithm. We provide many different parameterizations in order to check the robustness of the results. The interpretation of the tables is the same as those in the main text. The results presented here use the same instances and realization tables as those presented in the main text.}

\subsection*{Heavy-Tail MAB Instance}

\begin{table}[H]
\centering
\begin{tabular}{|c|c|c|c|}
\hline
   & $\TS$  & $\DEG$  & $\DG$ \\ \hline
$\TS$
    & \makecell{\textbf{0.054} $\pm$0.01}
    & \makecell{\textbf{0.16} $\pm$0.02}
    & \makecell{\textbf{0.18} $\pm$0.02} \\\hline
$\DEG$
    & \makecell{\textbf{0.33} $\pm$0.03}
    & \makecell{\textbf{0.31} $\pm$0.02}
    & \makecell{\textbf{0.26} $\pm$0.02} \\\hline
$\DG$
    & \makecell{\textbf{0.39} $\pm$0.03}
    & \makecell{\textbf{0.41} $\pm$0.03}
    & \makecell{\textbf{0.33} $\pm$0.02} \\\hline
\end{tabular}
\caption{Temporary Monopoly:  Heavy Tail, $X= 50$}
\vspace{-6mm}
\end{table}

\begin{table}[H]
\centering
\begin{tabular}{|c|c|c|c|}
\hline
   & $\TS$  & $\DEG$  & $\DG$ \\ \hline
$\TS$
    & \makecell{\textbf{0.003} $\pm$0.003}
    & \makecell{\textbf{0.083} $\pm$0.02}
    & \makecell{\textbf{0.17} $\pm$0.02} \\\hline
$\DEG$
    & \makecell{\textbf{0.045} $\pm$0.01}
    & \makecell{\textbf{0.25} $\pm$0.02}
    & \makecell{\textbf{0.23} $\pm$0.02} \\\hline
$\DG$
    & \makecell{\textbf{0.12} $\pm$0.02}
    & \makecell{\textbf{0.36} $\pm$0.03}
    & \makecell{\textbf{0.3} $\pm$0.02} \\\hline
\end{tabular}
\caption{Temporary Monopoly:  Heavy Tail, $X= 200$}
\vspace{-6mm}
\end{table}

\begin{table}[H]
\centering
\begin{tabular}{|c|c|c|c|}
\hline
   & $\TS$  & $\DEG$  & $\DG$ \\ \hline
$\TS$
    & \makecell{\textbf{0.0017} $\pm$0.002}
    & \makecell{\textbf{0.059} $\pm$0.01}
    & \makecell{\textbf{0.16} $\pm$0.02} \\\hline
$\DEG$
    & \makecell{\textbf{0.029} $\pm$0.007}
    & \makecell{\textbf{0.23} $\pm$0.02}
    & \makecell{\textbf{0.23} $\pm$0.02} \\\hline
$\DG$
    & \makecell{\textbf{0.097} $\pm$0.02}
    & \makecell{\textbf{0.34} $\pm$0.03}
    & \makecell{\textbf{0.29} $\pm$0.02} \\\hline
\end{tabular}
\caption{Temporary Monopoly:  Heavy Tail, $X= 300$}
\vspace{-6mm}
\end{table}

\begin{table}[H]
\centering
\begin{tabular}{|c|c|c|c|}
\hline
   & $\TS$  & $\DEG$  & $\DG$ \\ \hline
$\TS$
    & \makecell{\textbf{0.002} $\pm$0.003}
    & \makecell{\textbf{0.043} $\pm$0.01}
    & \makecell{\textbf{0.16} $\pm$0.02} \\\hline
$\DEG$
    & \makecell{\textbf{0.03} $\pm$0.007}
    & \makecell{\textbf{0.21} $\pm$0.02}
    & \makecell{\textbf{0.24} $\pm$0.02} \\\hline
$\DG$
    & \makecell{\textbf{0.091} $\pm$0.01}
    & \makecell{\textbf{0.32} $\pm$0.03}
    & \makecell{\textbf{0.3} $\pm$0.02} \\\hline
\end{tabular}
\caption{Temporary Monopoly:  Heavy Tail, $X= 500$}
\vspace{-6mm}
\end{table}

\subsection*{Needle-In-Haystack MAB Instance}

\begin{table}[H]
\centering
\begin{tabular}{|c|c|c|c|}
\hline
   & $\TS$  & $\DEG$  & $\DG$ \\ \hline
$\TS$
    & \makecell{\textbf{0.34} $\pm$0.03}
    & \makecell{\textbf{0.4} $\pm$0.03}
    & \makecell{\textbf{0.48} $\pm$0.03} \\\hline
$\DEG$
    & \makecell{\textbf{0.22} $\pm$0.02}
    & \makecell{\textbf{0.34} $\pm$0.03}
    & \makecell{\textbf{0.42} $\pm$0.03} \\\hline
$\DG$
    & \makecell{\textbf{0.18} $\pm$0.02}
    & \makecell{\textbf{0.28} $\pm$0.02}
    & \makecell{\textbf{0.37} $\pm$0.03} \\\hline
\end{tabular}
\caption{Temporary Monopoly:  Needle-in-Haystack, $X= 50$}
\vspace{-6mm}
\end{table}

\begin{table}[H]
\centering
\begin{tabular}{|c|c|c|c|}
\hline
   & $\TS$  & $\DEG$  & $\DG$ \\ \hline
$\TS$
    & \makecell{\textbf{0.17} $\pm$0.02}
    & \makecell{\textbf{0.31} $\pm$0.03}
    & \makecell{\textbf{0.41} $\pm$0.03} \\\hline
$\DEG$
    & \makecell{\textbf{0.13} $\pm$0.02}
    & \makecell{\textbf{0.26} $\pm$0.02}
    & \makecell{\textbf{0.36} $\pm$0.03} \\\hline
$\DG$
    & \makecell{\textbf{0.093} $\pm$0.02}
    & \makecell{\textbf{0.23} $\pm$0.02}
    & \makecell{\textbf{0.33} $\pm$0.03} \\\hline
\end{tabular}
\caption{Temporary Monopoly:  Needle-in-Haystack, $X= 200$}
\vspace{-6mm}
\end{table}

\begin{table}[H]
\centering
\begin{tabular}{|c|c|c|c|}
\hline
   & $\TS$  & $\DEG$  & $\DG$ \\ \hline
$\TS$
    & \makecell{\textbf{0.1} $\pm$0.02}
    & \makecell{\textbf{0.28} $\pm$0.03}
    & \makecell{\textbf{0.39} $\pm$0.03} \\\hline
$\DEG$
    & \makecell{\textbf{0.089} $\pm$0.02}
    & \makecell{\textbf{0.23} $\pm$0.02}
    & \makecell{\textbf{0.36} $\pm$0.03} \\\hline
$\DG$
    & \makecell{\textbf{0.05} $\pm$0.01}
    & \makecell{\textbf{0.21} $\pm$0.02}
    & \makecell{\textbf{0.33} $\pm$0.03} \\\hline
\end{tabular}
\caption{Temporary Monopoly:  Needle-in-Haystack, $X= 300$}
\vspace{-6mm}
\end{table}

\begin{table}[H]
\centering
\begin{tabular}{|c|c|c|c|}
\hline
   & $\TS$  & $\DEG$  & $\DG$ \\ \hline
$\TS$
    & \makecell{\textbf{0.053} $\pm$0.01}
    & \makecell{\textbf{0.23} $\pm$0.02}
    & \makecell{\textbf{0.37} $\pm$0.03} \\\hline
$\DEG$
    & \makecell{\textbf{0.051} $\pm$0.01}
    & \makecell{\textbf{0.2} $\pm$0.02}
    & \makecell{\textbf{0.33} $\pm$0.03} \\\hline
$\DG$
    & \makecell{\textbf{0.031} $\pm$0.009}
    & \makecell{\textbf{0.18} $\pm$0.02}
    & \makecell{\textbf{0.31} $\pm$0.02} \\\hline
\end{tabular}
\caption{Temporary Monopoly:  Needle-in-Haystack, $X= 500$}
\vspace{-6mm}
\end{table}

\subsection*{Uniform MAB Instance}

\begin{table}[H]
\centering
\begin{tabular}{|c|c|c|c|}
\hline
   & $\TS$  & $\DEG$  & $\DG$ \\ \hline
$\TS$
    & \makecell{\textbf{0.27} $\pm$0.03}
    & \makecell{\textbf{0.21} $\pm$0.02}
    & \makecell{\textbf{0.26} $\pm$0.02} \\\hline
$\DEG$
    & \makecell{\textbf{0.39} $\pm$0.03}
    & \makecell{\textbf{0.3} $\pm$0.03}
    & \makecell{\textbf{0.34} $\pm$0.03} \\\hline
$\DG$
    & \makecell{\textbf{0.39} $\pm$0.03}
    & \makecell{\textbf{0.31} $\pm$0.02}
    & \makecell{\textbf{0.33} $\pm$0.02} \\\hline
\end{tabular}
\caption{Temporary Monopoly:  Uniform, $X= 50$}
\vspace{-6mm}
\end{table}

\begin{table}[H]
\centering
\begin{tabular}{|c|c|c|c|}
\hline
   & $\TS$  & $\DEG$  & $\DG$ \\ \hline
$\TS$
    & \makecell{\textbf{0.12} $\pm$0.02}
    & \makecell{\textbf{0.16} $\pm$0.02}
    & \makecell{\textbf{0.2} $\pm$0.02} \\\hline
$\DEG$
    & \makecell{\textbf{0.25} $\pm$0.02}
    & \makecell{\textbf{0.24} $\pm$0.02}
    & \makecell{\textbf{0.29} $\pm$0.02} \\\hline
$\DG$
    & \makecell{\textbf{0.23} $\pm$0.02}
    & \makecell{\textbf{0.24} $\pm$0.02}
    & \makecell{\textbf{0.29} $\pm$0.02} \\\hline
\end{tabular}
\caption{Temporary Monopoly:  Uniform, $X= 200$}
\vspace{-6mm}
\end{table}

\begin{table}[H]
\centering
\begin{tabular}{|c|c|c|c|}
\hline
   & $\TS$  & $\DEG$  & $\DG$ \\ \hline
$\TS$
    & \makecell{\textbf{0.094} $\pm$0.02}
    & \makecell{\textbf{0.15} $\pm$0.02}
    & \makecell{\textbf{0.2} $\pm$0.02} \\\hline
$\DEG$
    & \makecell{\textbf{0.2} $\pm$0.02}
    & \makecell{\textbf{0.23} $\pm$0.02}
    & \makecell{\textbf{0.29} $\pm$0.02} \\\hline
$\DG$
    & \makecell{\textbf{0.21} $\pm$0.02}
    & \makecell{\textbf{0.23} $\pm$0.02}
    & \makecell{\textbf{0.29} $\pm$0.02} \\\hline
\end{tabular}
\caption{Temporary Monopoly:  Uniform, $X= 300$}
\vspace{-6mm}
\end{table}

\begin{table}[H]
\centering
\begin{tabular}{|c|c|c|c|}
\hline
   & $\TS$  & $\DEG$  & $\DG$ \\ \hline
$\TS$
    & \makecell{\textbf{0.061} $\pm$0.01}
    & \makecell{\textbf{0.12} $\pm$0.02}
    & \makecell{\textbf{0.2} $\pm$0.02} \\\hline
$\DEG$
    & \makecell{\textbf{0.17} $\pm$0.02}
    & \makecell{\textbf{0.21} $\pm$0.02}
    & \makecell{\textbf{0.29} $\pm$0.02} \\\hline
$\DG$
    & \makecell{\textbf{0.18} $\pm$0.02}
    & \makecell{\textbf{0.22} $\pm$0.02}
    & \makecell{\textbf{0.29} $\pm$0.02} \\\hline
\end{tabular}
\caption{Temporary Monopoly:  Uniform, $X= 500$}
\vspace{-6mm}
\end{table}

\section{Reputation vs. Data Advantage}

This section presents all experiments on data vs. reputation advantage (Section \ref{sec:barriers}).

Each experiment is presented as a table with the same semantics as in the main text. Namely, each cell in the table describes the duopoly game between the entrant's algorithm (the {\bf row}) and the incumbent's algorithm (the {\bf column}). The cell specifies the entrant's market share for the rounds in which hit was present: the average (in bold) and the 95\% confidence interval. NB: smaller average is better for the incumbent.

\begin{table}[H]
\centering
\begin{tabular}{|c|c|c|c|}
\hline
   & $\TS$  & $\DEG$  & $\DG$ \\ \hline
$\TS$
    & \makecell{\textbf{ 0.0096 } $\pm$ 0.006}
    & \makecell{\textbf{ 0.11 } $\pm$ 0.02}
    & \makecell{\textbf{ 0.18 } $\pm$ 0.02} \\\hline
$\DEG$
    & \makecell{\textbf{ 0.073 } $\pm$ 0.01}
    & \makecell{\textbf{ 0.29 } $\pm$ 0.02}
    & \makecell{\textbf{ 0.25 } $\pm$ 0.02} \\\hline
$\DG$
    & \makecell{\textbf{ 0.15 } $\pm$ 0.02}
    & \makecell{\textbf{ 0.39 } $\pm$ 0.03}
    & \makecell{\textbf{ 0.33 } $\pm$ 0.02} \\\hline
\end{tabular}
\caption{Data Advantage: Heavy Tail, $X = 200$}
\vspace{-6mm}
\end{table}

\begin{table}[H]
\centering
\begin{tabular}{|c|c|c|c|}
\hline
   & $\TS$  & $\DEG$  & $\DG$ \\ \hline
$\TS$
    & \makecell{\textbf{ 0.021 } $\pm$ 0.009}
    & \makecell{\textbf{ 0.16 } $\pm$ 0.02}
    & \makecell{\textbf{ 0.21 } $\pm$ 0.02} \\\hline
$\DEG$
    & \makecell{\textbf{ 0.26 } $\pm$ 0.03}
    & \makecell{\textbf{ 0.3 } $\pm$ 0.02}
    & \makecell{\textbf{ 0.26 } $\pm$ 0.02} \\\hline
$\DG$
    & \makecell{\textbf{ 0.34 } $\pm$ 0.03}
    & \makecell{\textbf{ 0.4 } $\pm$ 0.03 }
    & \makecell{\textbf{ 0.33 } $\pm$ 0.02} \\\hline
\end{tabular}
\caption{Reputation Advantage: Heavy Tail, $X=200$}
\vspace{-6mm}
\end{table}

\begin{table}[H]
\centering
\begin{tabular}{|c|c|c|c|}
\hline
   & $\TS$  & $\DEG$  & $\DG$ \\ \hline
$\TS$
    & \makecell{\textbf{ 0.25 } $\pm$ 0.03}
    & \makecell{\textbf{ 0.36 } $\pm$ 0.03}
    & \makecell{\textbf{ 0.45 } $\pm$ 0.03} \\\hline
$\DEG$
    & \makecell{\textbf{ 0.21 } $\pm$ 0.02}
    & \makecell{\textbf{ 0.32 } $\pm$ 0.03}
    & \makecell{\textbf{ 0.41 } $\pm$ 0.03} \\\hline
$\DG$
    & \makecell{\textbf{ 0.18 } $\pm$ 0.02}
    & \makecell{\textbf{ 0.29 } $\pm$ 0.03}
    & \makecell{\textbf{ 0.4 } $\pm$ 0.03} \\\hline
\end{tabular}
\caption{Data Advantage: Needle-in-Haystack, $X=200$}
\vspace{-6mm}
\end{table}

\begin{table}[H]
\centering
\begin{tabular}{|c|c|c|c|}
\hline
   & $\TS$  & $\DEG$  & $\DG$ \\ \hline
$\TS$
    & \makecell{\textbf{ 0.35 } $\pm$ 0.03}
    & \makecell{\textbf{ 0.43 } $\pm$ 0.03}
    & \makecell{\textbf{ 0.52 } $\pm$ 0.03} \\\hline
$\DEG$
    & \makecell{\textbf{ 0.26 } $\pm$ 0.03 }
    & \makecell{\textbf{ 0.36 } $\pm$ 0.03}
    & \makecell{\textbf{ 0.43 } $\pm$ 0.03} \\\hline
$\DG$
    & \makecell{\textbf{ 0.19 } $\pm$ 0.02}
    & \makecell{\textbf{ 0.3 } $\pm$ 0.02}
    & \makecell{\textbf{ 0.36 } $\pm$ 0.02} \\\hline
\end{tabular}
\caption{Reputation Advantage: Needle-in-Haystack, $X=200$}
\vspace{-6mm}
\end{table}

\begin{table}[H]
\centering
\begin{tabular}{|c|c|c|c|}
\hline
   & $\TS$  & $\DEG$  & $\DG$ \\ \hline
$\TS$
    & \makecell{\textbf{ 0.27 } $\pm$ 0.03}
    & \makecell{\textbf{ 0.23 } $\pm$ 0.02}
    & \makecell{\textbf{ 0.27 } $\pm$ 0.02} \\\hline
$\DEG$
    & \makecell{\textbf{ 0.4 } $\pm$ 0.03}
    & \makecell{\textbf{ 0.3 } $\pm$ 0.02 }
    & \makecell{\textbf{ 0.32 } $\pm$ 0.02} \\\hline
$\DG$
    & \makecell{\textbf{ 0.36 } $\pm$ 0.03}
    & \makecell{\textbf{ 0.29 } $\pm$ 0.02}
    & \makecell{\textbf{ 0.3 } $\pm$ 0.02} \\\hline
\end{tabular}
\caption{Reputation Advantage: Uniform, $X=200$}
\vspace{-6mm}
\end{table}

\begin{table}[H]
\centering
\begin{tabular}{|c|c|c|c|}
\hline
   & $\TS$  & $\DEG$  & $\DG$ \\ \hline
$\TS$
    & \makecell{\textbf{ 0.2 } $\pm$ 0.02}
    & \makecell{\textbf{ 0.22 } $\pm$ 0.02}
    & \makecell{\textbf{ 0.27 } $\pm$ 0.03} \\\hline
$\DEG$
    & \makecell{\textbf{ 0.33 } $\pm$ 0.03}
    & \makecell{\textbf{ 0.32 } $\pm$ 0.03}
    & \makecell{\textbf{ 0.35 } $\pm$ 0.03} \\\hline
$\DG$
    & \makecell{\textbf{ 0.32 } $\pm$ 0.03}
    & \makecell{\textbf{ 0.31 } $\pm$ 0.03}
    & \makecell{\textbf{ 0.35 } $\pm$ 0.03} \\\hline
\end{tabular}
\caption{Data Advantage: Uniform, $X=200$}
\vspace{-6mm}
\end{table}

\begin{table}[H]
\centering
\begin{tabular}{|c|c|c|c|}
\hline
   & $\TS$  & $\DEG$  & $\DG$ \\ \hline
$\TS$
    & \makecell{\textbf{0.0017} $\pm$0.002}
    & \makecell{\textbf{0.06} $\pm$0.01}
    & \makecell{\textbf{0.18} $\pm$0.02} \\\hline
$\DEG$
    & \makecell{\textbf{0.04} $\pm$0.009}
    & \makecell{\textbf{0.24} $\pm$0.02}
    & \makecell{\textbf{0.25} $\pm$0.02} \\\hline
$\DG$
    & \makecell{\textbf{0.12} $\pm$0.02}
    & \makecell{\textbf{0.35} $\pm$0.03}
    & \makecell{\textbf{0.33} $\pm$0.02} \\\hline
\end{tabular}
\caption{Data Advantage: Heavy-Tail, $X=500$}
\vspace{-6mm}
\end{table}

\begin{table}[H]
\centering
\begin{tabular}{|c|c|c|c|}
\hline
   & $\TS$  & $\DEG$  & $\DG$ \\ \hline
$\TS$
    & \makecell{\textbf{0.022} $\pm$0.009}
    & \makecell{\textbf{0.13} $\pm$0.02}
    & \makecell{\textbf{0.21} $\pm$0.02} \\\hline
$\DEG$
    & \makecell{\textbf{0.26} $\pm$0.03}
    & \makecell{\textbf{0.29} $\pm$0.02}
    & \makecell{\textbf{0.28} $\pm$0.02} \\\hline
$\DG$
    & \makecell{\textbf{0.33} $\pm$0.03}
    & \makecell{\textbf{0.39} $\pm$0.03}
    & \makecell{\textbf{0.34} $\pm$0.02} \\\hline
\end{tabular}
\caption{Reputation Advantage: Heavy-Tail, $X=500$}
\vspace{-6mm}
\end{table}

\begin{table}[H]
\centering
\begin{tabular}{|c|c|c|c|}
\hline
   & $\TS$  & $\DEG$  & $\DG$ \\ \hline
$\TS$
    & \makecell{\textbf{0.098} $\pm$0.02}
    & \makecell{\textbf{0.27} $\pm$0.03}
    & \makecell{\textbf{0.41} $\pm$0.03} \\\hline
$\DEG$
    & \makecell{\textbf{0.093} $\pm$0.02}
    & \makecell{\textbf{0.24} $\pm$0.02}
    & \makecell{\textbf{0.38} $\pm$0.03} \\\hline
$\DG$
    & \makecell{\textbf{0.064} $\pm$0.01}
    & \makecell{\textbf{0.22} $\pm$0.02}
    & \makecell{\textbf{0.37} $\pm$0.03} \\\hline
\end{tabular}
\caption{Data Advantage: Needle-in-Haystack, $X=500$}
\vspace{-6mm}
\end{table}

\begin{table}[H]
\centering
\begin{tabular}{|c|c|c|c|}
\hline
   & $\TS$  & $\DEG$  & $\DG$ \\ \hline
$\TS$
    & \makecell{\textbf{0.29} $\pm$0.03}
    & \makecell{\textbf{0.44} $\pm$0.03}
    & \makecell{\textbf{0.52} $\pm$0.03} \\\hline
$\DEG$
    & \makecell{\textbf{0.19} $\pm$0.02}
    & \makecell{\textbf{0.35} $\pm$0.03}
    & \makecell{\textbf{0.42} $\pm$0.03} \\\hline
$\DG$
    & \makecell{\textbf{0.15} $\pm$0.02}
    & \makecell{\textbf{0.27} $\pm$0.02}
    & \makecell{\textbf{0.35} $\pm$0.02} \\\hline
\end{tabular}
\caption{Reputation Advantage: Needle-in-Haystack, $X=500$}
\vspace{-6mm}
\end{table}

\begin{table}[H]
\centering
\begin{tabular}{|c|c|c|c|}
\hline
   & $\TS$  & $\DEG$  & $\DG$ \\ \hline
$\TS$
    & \makecell{\textbf{0.14} $\pm$0.02}
    & \makecell{\textbf{0.18} $\pm$0.02}
    & \makecell{\textbf{0.26} $\pm$0.03} \\\hline
$\DEG$
    & \makecell{\textbf{0.26} $\pm$0.02}
    & \makecell{\textbf{0.26} $\pm$0.02}
    & \makecell{\textbf{0.34} $\pm$0.03} \\\hline
$\DG$
    & \makecell{\textbf{0.25} $\pm$0.02}
    & \makecell{\textbf{0.27} $\pm$0.02}
    & \makecell{\textbf{0.34} $\pm$0.03} \\\hline
\end{tabular}
\caption{Data Advantage: Uniform, $X=500$}
\vspace{-6mm}
\end{table}

\begin{table}[H]
\centering
\begin{tabular}{|c|c|c|c|}
\hline
   & $\TS$  & $\DEG$  & $\DG$ \\ \hline
$\TS$
    & \makecell{\textbf{0.24} $\pm$0.02}
    & \makecell{\textbf{0.2} $\pm$0.02}
    & \makecell{\textbf{0.26} $\pm$0.02} \\\hline
$\DEG$
    & \makecell{\textbf{0.37} $\pm$0.03}
    & \makecell{\textbf{0.29} $\pm$0.02}
    & \makecell{\textbf{0.31} $\pm$0.02} \\\hline
$\DG$
    & \makecell{\textbf{0.35} $\pm$0.03}
    & \makecell{\textbf{0.27} $\pm$0.02}
    & \makecell{\textbf{0.3} $\pm$0.02} \\\hline
\end{tabular}
\caption{Reputation Advantage: Uniform, $X=500$}
\vspace{-6mm}
\end{table}

\section{Mean Reputation vs. Relative Reputation}

We present the experiments omitted from Section \ref{sec:revisited}. Namely, experiments on the Heavy-Tail MAB instance with $K=3$ arms, both for ``performance in isolation" and the permanent duopoly game. We find that $\DEG > \DG$ according to the mean reputation trajectory but that $\DG > \DEG$ according to the relative reputation trajectory \emph{and} in the competition game. As discussed in Section \ref{sec:revisited}, the same results also hold for $K = 10$ for the warm starts that we consider.

The result of the permanent duopoly experiment for this instance  is shown in Table \ref{ht_k3}.

\begin{table}[h]
\centering
\begin{tabular}{|c|c|c|c|}
  \hline
  & \multicolumn{3}{c|}{Heavy-Tail} \\
\hline
   & $T_0$ = 20 & $T_0$ = 250 & $T_0$ = 500 \\ \hline
\TS vs. \DG
  & \makecell{\textbf{0.4} $\pm$0.02\\ \Eeog 770 (0)}
    & \makecell{\textbf{0.59} $\pm$0.01\\ \Eeog 2700 (2979.5)}
    & \makecell{\textbf{0.6} $\pm$0.01\\ \Eeog 2700 (3018)} \\ \hline
\TS vs. \DEG
    & \makecell{\textbf{0.46} $\pm$0.02 \\ \Eeog 830 (0)}
    & \makecell{\textbf{0.73} $\pm$0.01 \\ \Eeog 2500 (2576.5)}
    & \makecell{\textbf{0.72} $\pm$0.01 \\ \Eeog 2700 (2862)} \\ \hline
\DG vs. \DEG
    & \makecell{\textbf{0.61} $\pm$0.01 \\ \Eeog 1400 (556)}
    & \makecell{\textbf{0.61} $\pm$0.01 \\ \Eeog 2400 (2538.5)}
    & \makecell{\textbf{0.6} $\pm$0.01 \\ \Eeog 2400 (2587.5)} \\\hline
\end{tabular}
\caption{Duopoly Experiment: Heavy-Tail, $K=3$, $T=5000$.\\
Each cell describes a game between two algorithms, call them Alg1 vs. Alg2, for a particular value of the warm start $T_0$. Line 1 in the cell is the market share of Alg 1: the average (in bold) and the 95\% confidence band.
%For example, the cell in the top left indicates that TS gets on average 64\% of the market when played against DG.
Line 2 specifies the ``effective end of game" (\Eeog): the average and the median (in brackets). }
\label{ht_k3}
\end{table}

The mean reputation trajectories for algorithms' performance in isolation:
\begin{center}
\includegraphics[scale=0.35]{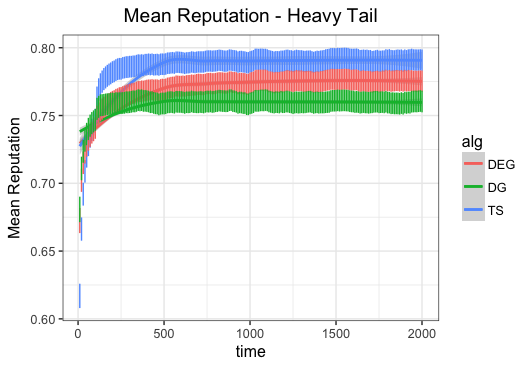} \\
\end{center}

Finally, the relative reputation trajectory of \DEG vs. \DG:
\begin{center}
\includegraphics[scale=0.35]{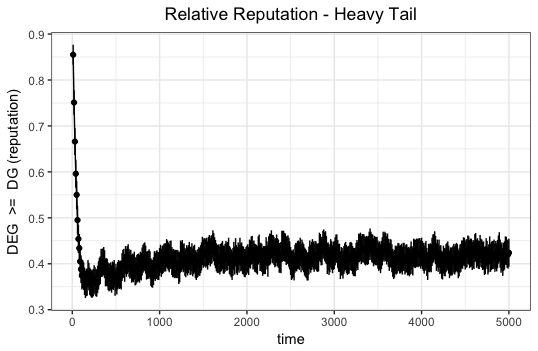}
\end{center}
\end{appendices}

\end{document}